\documentclass[pra,aps,twocolumn,footinbib,superscriptaddress,citeautoscript,longbibliography]{revtex4-1}

\usepackage{amssymb}
\usepackage{amsmath}
\usepackage{amsfonts}
\usepackage{color}
\usepackage{graphicx}
\usepackage{bm}
\usepackage{todonotes}

\usepackage[unicode]{hyperref}
\hypersetup{
   unicode=true,          
   plainpages=false,
   colorlinks=true,       
   citecolor=blue,        
}

\newcommand* {\vek}[1]{{\ensuremath{\bm{\mathrm{#1}}}}}

\newcommand* {\kk}{\vek{k}}

\newcommand* {\ee}{\ensuremath{\mathrm{e}}}

\graphicspath{{.}{./FIG/}}

\begin{document}

\title{Chiral twodimensional \textit{p}-wave superfluid from
\textit{s}-wave pairing in the BEC regime}

\author{K. Thompson}
\affiliation{School of Chemical and Physical Sciences, Victoria
University of Wellington, PO Box 600, Wellington 6140, New Zealand}
\affiliation{Dodd-Walls Centre for Photonic and Quantum Technologies,
PO Box 56, Dunedin 9056, New Zealand}

\author{J. Brand}
\affiliation{Centre for Theoretical Chemistry and Physics, and New
Zealand Institute for Advanced Study, Massey University, Private
Bag 102904 NSMC, Auckland 0745, New Zealand}
\affiliation{Dodd-Walls Centre for Photonic and Quantum Technologies,
PO Box 56, Dunedin 9056, New Zealand}

\author{U. Z\"ulicke}
\email{uli.zuelicke@vuw.ac.nz}
\affiliation{School of Chemical and Physical Sciences, Victoria
University of Wellington, PO Box 600, Wellington 6140, New Zealand}
\affiliation{Dodd-Walls Centre for Photonic and Quantum Technologies,
PO Box 56, Dunedin 9056, New Zealand}
\affiliation{Department of Physics, University of Basel,
Klingelbergstrasse 82, CH-4056 Basel, Switzerland}

\date{\today}

\begin{abstract}

Twodimensional spin-orbit-coupled Fermi gases subject to
\textit{s}-wave pairing can be driven into a topological phase by
increasing the Zeeman spin splitting beyond a critical value. In the
topological phase, the system exhibits the hallmarks of chiral
\textit{p}-wave superfluidity, including exotic Majorana excitations.
Previous theoretical studies of this realization of a twodimensional
topological Fermi superfluid have focused on the BCS regime where the
\textit{s}-wave Cooper pairs are only weakly bound and, hence, the
induced chiral \textit{p}-wave order parameter has a small magnitude.
Motivated by the goal to identify potential new ways for the
experimental realization of robust topological superfluids in
ultra-cold atom gases, we study the BCS-to-BEC crossover driven by
increasing the Cooper-pair binding energy for this system. In
particular, we obtain phase diagrams in the parameter space of
two-particle bound-state energy and Zeeman spin-splitting energy.
Ordinary characteristics of the BCS-to-BEC crossover, in particular
the shrinking and eventual disappearance of the Fermi surface, are
observed in the nontopological phase. In contrast, the topological
phase retains all features of chiral \textit{p}-wave superfluidity,
including a well-defined underlying Fermi surface, even for large
\textit{s}-wave pair-binding energies. Compared to the BCS limit, the
topological superfluid in the BEC regime turns out to be better
realizable even for only moderate magnitude of spin-orbit coupling
because the chiral \textit{p}-wave order parameter is generally larger
and remnants of \textit{s}-wave pairing are suppressed. We identify
optimal parameter ranges that can aid further experimental
investigations and elucidate the underlying physical reason for the
persistence of the chiral \textit{p}-wave superfluid.

\end{abstract}

\maketitle

\section{Introduction and overview of main results}

One of the earliest proposed pathways towards realization of a
twodimensional (2D) topological superfluid (TSF)~\cite{Sato2017} is
based on \textit{s}-wave pairing of spin-$\frac{1}{2}$ fermions
subject to spin-orbit coupling and Zeeman spin splitting~\cite{Fu2008,
Zhang2008,Sau2010,Alicea2010,Sato2010}. In the absence of spin-orbit
coupling,  a population imbalance in the spin components (equivalent
to nonzero Zeeman splitting) tends to destroy \textit{s}-wave
superfluidity due to the mismatch of the spin-$\uparrow$ and
spin-$\downarrow$ Fermi surfaces for weak-coupling superfluids
\cite{Chandrasekhar1962,Clogston1962}. With strong \textit{s}-wave
attraction, phase separation between superfluid and normal phases
ensues in this case \cite{He2008}. Adding 2D spin-orbit coupling
(e.g., of Rashba form~\cite{Bychkov1984,Winkler2003,Galitski2013})
permits a pairing instability even for unmatched Fermi surfaces and
re-establishes a homogeneous superfluid ground state with gapped
fermionic quasiparticle excitations. The pairing field for each spin
component separately~\cite{Brand2018} now obtains the characteristics
of a chiral 2D \textit{p}-wave superfluid~\cite{Kallin2016}.
Increasing the Zeeman coupling energy $h$ beyond the critical value
\begin{equation}\label{eq:critZeem}
h_\mathrm{c} = \sqrt{\mu^2 + |\Delta|^2}
\end{equation}
quenches one of the Fermi surfaces, and the system enters a TSF
phase. In Eq.~(\ref{eq:critZeem}), $\Delta$ and $\mu$ denote the
selfconsistent \textit{s}-wave pair potential and chemical potential,
respectively. Bearing all the characteristics of a 2D spinless
\textit{p}-wave superfluid, a nontrivial topological invariant can be
defined~\cite{Sato2017}, and Majorana quasiparticle
excitations are present at boundaries~\cite{Jackiw1976,Fu2008} and in
vortex cores~\cite{Kopnin1991,Volovik1999,Read2000,Ivanov2001,
Gurarie2007} by virtue of an index theorem \cite{Tewari2007a}.
Majorana zero modes are considered promising candidates for enabling
fault-tolerant quantum-information processing~\cite{DasSarma2015}.

Intense efforts towards experimental implementation of 2D TSFs using
the above-described route have so far been thwarted by the deleterious
effect of Zeeman-splitting-inducing magnetic fields on
superconductivity in typical materials~\cite{Loder2015}, as well as
basic physical constraints on the magnitude of spin-orbit coupling
reachable in solids~\cite{Winkler2003} and ultra-cold atom
gases~\cite{Zhai2015,Zhang2018}. Our present study shows that a
possible way around the latter limitation would be to access the
strong-coupling regime of the \textit{s}-wave pairing, which is
commonly referred to as the BEC regime~\cite{Leggett1980a,
Leggett1980b,Randeria1990,Parish2015,Strinati2018}.

\begin{figure}[t]
\includegraphics[width=0.85\columnwidth]{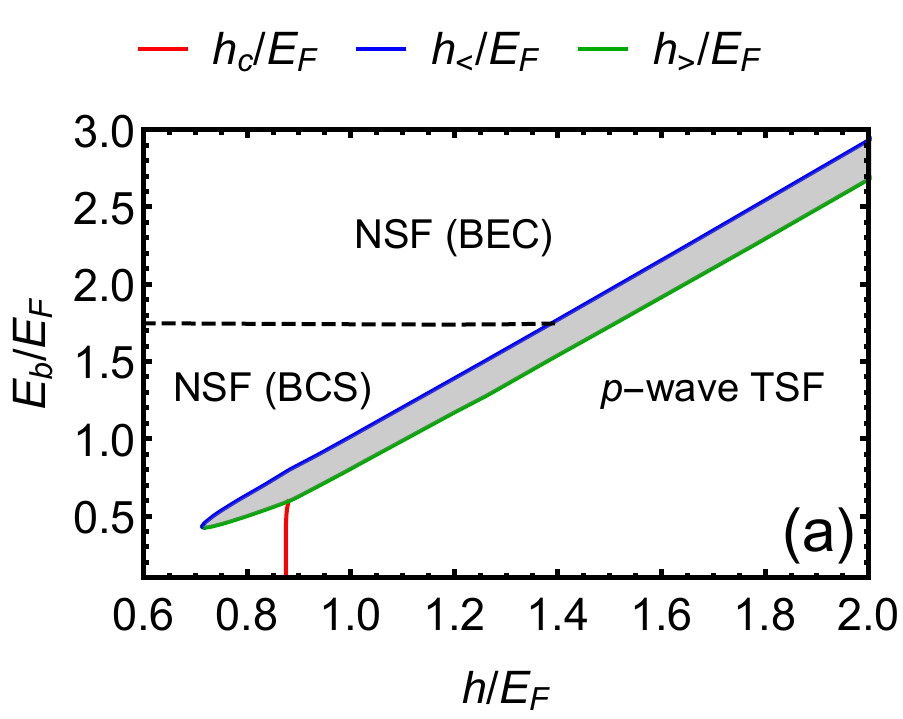}\\[0.2cm]
\includegraphics[width=0.85\columnwidth]{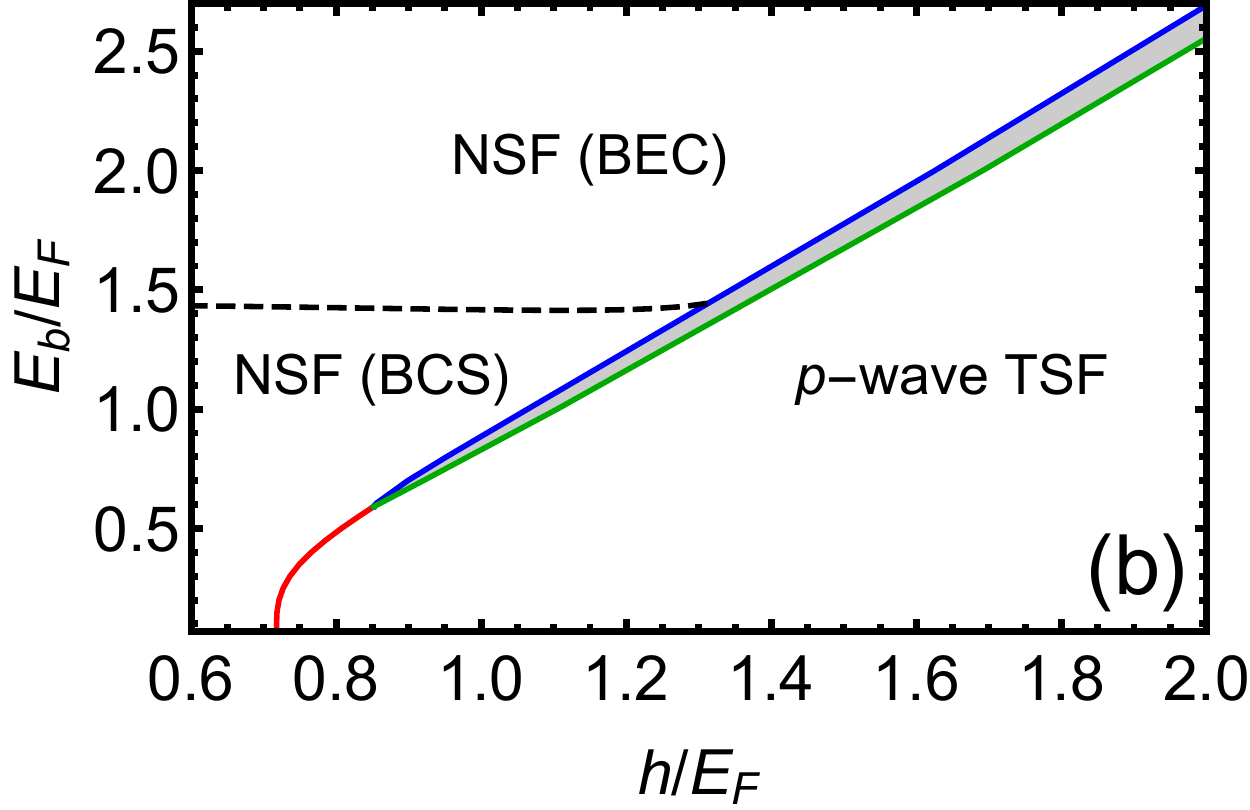}
\caption{\label{fig:phaseDia}%
Zero-temperature mean-field phase diagrams, in the parameter space of
two-particle \textit{s}-wave bound-state energy $E_\mathrm{b}$ and
Zeeman energy $h$, for a spin-orbit-coupled 2D Fermi gas with fixed
density $n = m E_\mathrm{F}/(\pi\hbar^2)\equiv k_\mathrm{F}^2/(2\pi)$.
Panel (a) [(b)] depicts the case where the dimensionless parameter
$\lambda k_\mathrm{F}/E_\mathrm{F}$ measuring the spin-orbit-coupling
strength equals $0.50$ [$0.75$], which illustrates a
small-(large-)spin-orbit-coupling situation. The shaded region $h_< <
h < h_>$, with $h_<$ ($h_>$) indicated by the blue (green) line, is in
the phase-separated first-order-transition regime that emerges for
$E_\mathrm{b}>E_\mathrm{b}^{(\mathrm{c})}$ and $h > h^{(\mathrm{c})}$.
The critical Zeeman energy $h_\mathrm{c}$, defined via
Eq.~(\ref{eq:critZeem}) and indicated by the red curve, delineates the
second-order transition between an ordinary nontopological superfluid
(NSF) and a topological superfluid (TSF). From the point when the
$h_\mathrm{c}\big(E_\mathrm{b}\big)$-curve reaches the region where
phase separation occurs, the topological transition is switched from
second to first order. The BCS-to-BEC-crossover boundary (dashed line)
has been determined via the condition $\mu(E_\mathrm{b},h)=0$.}
\end{figure}

The main insights reached in our work are underpinned by
zero-temperature phase diagrams in the parameter space of two-particle
\textit{s}-wave bound-state energy $E_\mathrm{b}$ \cite{noteEb} and
Zeeman energy $h$ as illustrated in Fig.~\ref{fig:phaseDia}. These
show a second-order transition line (red) between the nontopologial
and topological superfluids at small $E_\mathrm{b}$ being replaced by
a first-order phase transition at larger $E_\mathrm{b}$. In the phase
diagram at constant particle density $n$, enforced by measuring
energies in terms of the Fermi energy $E_\mathrm{F} = \pi\hbar^2 n/m$,
the first-order phase transition manifests itself as a region without
a uniform-density ground state (grey) where phase separation into
spatially separated domains ensues. Critical Zeeman-energy values
$h_<$ (blue) and $h_>$ (green) delimit the phase-separation region at
fixed $E_\mathrm{b}$, defining two curves in the phase diagram. The
properties of the phase-separation region itself have been the subject
of previous work~\cite{Yi2011,Yang2012}, and we also provide a few
more details later on. However, the main focus of our present study is
the careful determination of the location of the boundaries $h_<$ and
$h_>$ and the exploration of the adjacent homogeneous phases,
especially in the regime where $E_\mathrm{b}/E_\mathrm{F} \gtrsim 1$.

Comparison of our results with those obtained previously for
spin-imbalanced 2D Fermi superfluids~\footnote{See, e.g., Fig.~2(a) in
Ref.~\cite{He2008}.} helps to elucidate the physical consequences of
finite spin-orbit coupling. The magnitude of the latter is most
conveniently measured in terms of the dimensionless parameter $\lambda
k_\mathrm{F}/E_\mathrm{F}$ that also involves the density-dependent
Fermi wave number $k_\mathrm{F}\equiv \sqrt{2\pi n}$. One important
effect of finite $\lambda$ is to shift the low-$E_\mathrm{b}$ boundary
of the phase-separation region from zero to finite values of
$E_\mathrm{b}$~\cite{He2008}, and a second effect is to establish the
TSF phase~\cite{Yang2012} for sufficiently high Zeeman energy
$h>\mathrm{max}\{h_c, h_>\}$ in place of the fully polarized normal
phase found for $\lambda=0$~\cite{He2008}.

Our present study shows that the character of the TSF phase emerging
in the BEC regime of the underpinning \textit{s}-wave pairing
($E_\mathrm{b}/E_\mathrm{F}\gtrsim 1$) is fundamentally similar to the
TSF occurring in the BCS limit ($E_\mathrm{b}/E_\mathrm{F}\ll 1$). In
particular, for the entire TSF region in the phase diagram, the system
exhibits canonical signatures of an underlying Fermi surface. As
discussed by Sensarma \emph{et al.}~\cite{Sensarma2007}, an underlying
Fermi surface can be robustly defined even in strong-coupling
fermionic superfluids by a number of alternative definitions, such as
a zero crossing of the single-particle Greens function
$G(\mathbf{k},0)$, a drop in the single-particle momentum distribution
$n(\mathbf{k})$, or, if available, by a minimum of the quasiparticle
dispersion relation. Our results for the TSF are in stark contrast to
the nontopological superfluid (NSF) phase where the Fermi surface
shrinks and eventually disappears as $E_\mathrm{b}/E_\mathrm{F}$
increases and the BEC regime is entered. This is expected from the
known phenomenology of the BCS-to-BEC crossover for \textit{s}-wave
pairing~\cite{Nozieres1985,Chen2005,Astrakharchik2005} and illustrated
by recent Quantum-Monte-Carlo results~\cite{Shi2015,Shi2016}.

\begin{figure}[t]
\includegraphics[width=0.85\columnwidth]{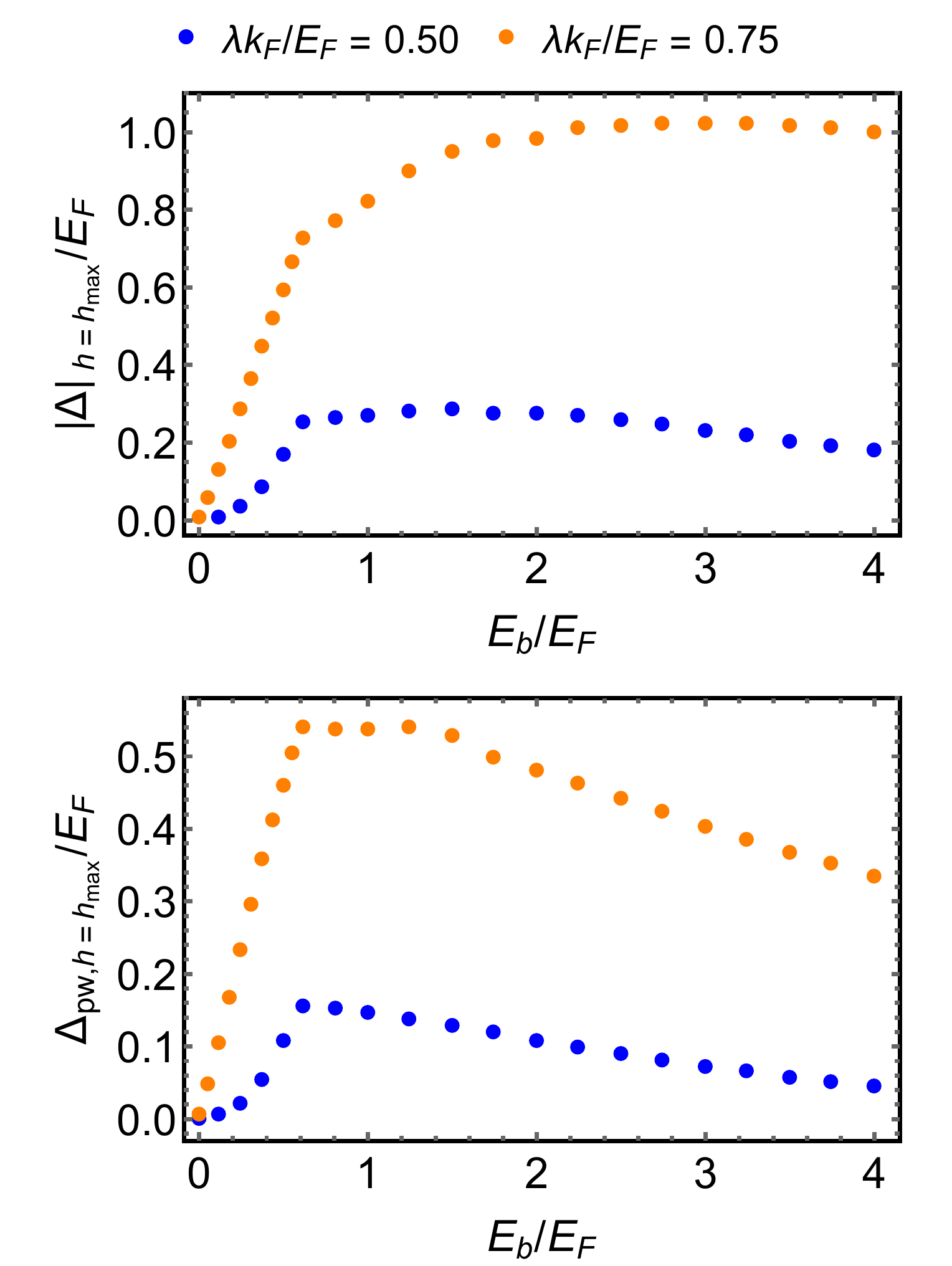}
\caption{\label{fig:maxDelta}%
The magnitude of the \textit{s}-wave pair potential $\Delta$ in the
homogeneous topological-superfluid phase of a spin-orbit-coupled 2D
Fermi gas with fixed density $n = m E_\mathrm{F}/(\pi\hbar^2)\equiv
k_\mathrm{F}^2/(2\pi)$ is maximized, for fixed two-particle
bound-state energy $E_\mathrm{b}$ and spin-orbit-coupling strength
$\lambda$, right after the transition at Zeeman energy $h_{\text{max}}
=\mathrm{max}\{h_\mathrm{c}, h_>\}$ (see Fig.~\ref{fig:phaseDia}).
Panel (a) shows the dependence of this maximum value, $|\Delta|_{h=
h_{\text{max}}}$, on $E_\mathrm{b}$ for two fixed values of the
dimensionless spin-orbit-coupling strength $\lambda k_\mathrm{F}/
E_\mathrm{F}$. Panel (b) is a plot of the emergent
chiral-\textit{p}-wave gap $\Delta_\mathrm{pw}$, extracted from the
low-energy quasiparticle dispersion, for the same parameters.}
\end{figure}

The defining element of the 2D TSF is an emergent chiral
\textit{p}-wave order parameter $\Delta_\mathrm{pw}$ whose magnitude
provides the energy scale of the quasiparticle-excitation gap. Its
value is proportional to the spin-orbit-coupling strength $\lambda$
and the modulus $|\Delta|$ of the \textit{s}-wave pair potential, but
inversely related to the spin-splitting (Zeeman) energy scale
$h$~\cite{Zhang2008,Alicea2010,Sato2010,Seo2012,Brand2018}. Given that
increasing $\lambda$ has adverse side effects such as heating of the
atom gas~\cite{Zhai2015} in currently available experimental schemes,
maximizing $\Delta_\mathrm{pw}$ needs to be pursued by other means. As
$|\Delta|/E_\mathrm{F}$ is a monotonously increasing function of
$E_\mathrm{b}/E_\mathrm{F}$ but is suppressed with increasing
$h/E_\mathrm{F}$ (see, e.g., Refs.~\cite{He2013,Brand2018} and below),
its practically largest magnitude occurs just after the transition to
the homogeneous TSF phase at $h_{\text{max}}=\mathrm{max}\{
h_\mathrm{c}, h_>\}$. Figure~\ref{fig:maxDelta}(a) illustrates the
dependence of this value, $|\Delta|_{h=h_{\text{max}}}$, on both
$E_\mathrm{b}$ and $\lambda$. It reveals a maximum that gets broader
and larger as the parameter $\lambda k_\mathrm{F}/E_\mathrm{F}$
increases. As the maximum value of $|\Delta|$ reaches values up to
$\sim\! E_\mathrm{F}$ typically, even for only moderately high values
of the spin-orbit-coupling strength, the TSF realized in the BEC
regime of \textit{s}-wave pairing presents a much more favorable
platform for useful study and application than would be available in
the BCS regime at the same value of $\lambda$. This is established
even more directly by measuring $\Delta_\mathrm{pw}$ as the gap in the
low-energy quasiparticle dispersion for the TSF. The values for
$\Delta_\mathrm{pw}$ found for the same parameter combinations that
maximize $|\Delta|$ are shown in Fig.~\ref{fig:maxDelta}(b). For both
fixed values of $\lambda k_\mathrm{F}/E_\mathrm{F}$, a maximum of
$\Delta_\mathrm{pw}$ occurs for $E_\mathrm{b}/E_\mathrm{F} \sim 1$,
followed by a broad range for which $\Delta_\mathrm{pw}$ is slowly
decreasing.

All quantitative results in this work were obtained within mean-field
theory, even though its validity for a 2D gas, in particular outside
of the weakly interacting regime, may not be taken for granted. We
nevertheless expect the qualitative physics, and in particular the
presence of a Fermi surface in the TSF phase, to be robust because the
topological property puts strong constraints on the many-body system.
We comment further on the physical reasons below. Mean-field
approximations have previously been found to provide useful insight
into zero-temperature phases, even when interactions are
strong~\cite{Yi2006,Parish2007,He2008,Fischer2013}. Quantitatively
more accurate predictions, in particular for finite temperature,
require more sophisticated approaches~\cite{Kuchiev1996,Yi2006,
Parish2007,Bertaina2011,Salasnich2015,He2015,Turlapov2017,Hu2018}.
The expected effects of beyond-mean-field corrections (quantum
fluctuations) is to suppress pairing gaps compared to mean-field
theory in the strongly interacting regime~\cite{He2015,Hu2018}. This
fact reinforces the optimal value $E_\mathrm{b}/E_\mathrm{F} \sim 1$
for realizing a robust TSF, as for $E_\mathrm{b}/E_\mathrm{F}\gg 1$,
the true value for the \textit{s}-wave pairing gap, and therefore also
$\Delta_\mathrm{pw}$, are likely to be much smaller than mean-field
theory predicts.

The physical reasons for the remarkable BCS-like behavior of the TSF
even when interactions are strong enough to place \textit{s}-wave
pairs into the BEC regime may be seen from a careful analysis of the
relevant low-energy part of the quasiparticle spectrum. A projection
of the mean-field equations to the majority-spin
component~\cite{Brand2018} yields a useful approximate expression for
the excitation gap and TSF order parameter
\begin{equation}\label{eq:pGap}
\Delta_\mathrm{pw} \approx |\Delta|\, \frac{\lambda\,
k_\mathrm{FS}}{h_{k_\mathrm{FS}}}\quad ,
\end{equation}
where $h_k = (h + \sqrt{h^2 + \lambda^2 k^2})/2$, and $k_\mathrm{FS}
\le \sqrt{2} k_\mathrm{F}$ is the radius of the Fermi surface in the
TSF phase (see end of Sec.~\ref{sec:results}). The projective
approximation is valid when $\Delta_\mathrm{pw}$ is small compared to
the Fermi energy $E_\mathrm{F}$, but this condition will be fulfilled
when spin-orbit coupling is not too strong, $\lambda k_\mathrm{F} <
E_\mathrm{F}$, in the TSF regime where $h>|\Delta|$ due to
Eq.~\eqref{eq:critZeem}. Note that this means that
$\Delta_\mathrm{pw}$ is bounded while the binding energy
$E_\mathrm{b}$ may be much larger. Within the same projective
approximation~\cite{Brand2018} and for $\lambda k_\mathrm{F} < h$, one
also obtains the estimate
\begin{equation}\label{kFSapprox}
k_\mathrm{FS} \approx k_\mathrm{F} \left[ \frac{\mu + h -
\frac{|\Delta|^2}{2 h}}{E_\mathrm{F} \left( 1 - \frac{1}{2}
\frac{\lambda^2 k_\mathrm{F}^2}{E_\mathrm{F} h} \right)}
\right]^{\frac{1}{2}} \quad ,
\end{equation}
which shows that the Fermi surface radius is finite, $k_\mathrm{FS} >
0$, as long as the Zeeman energy $h$ is sufficiently large. Thus the
large magnitude of the Zeeman energy required to reach the uniform TSF
phase ultimately ensures the persistence of BCS-like character of
chiral \textit{p}-wave pairing, even as the \textit{s}-wave
interaction is deep in the BEC regime. While the situation becomes
slightly more complex for very large spin-orbit-coupling strengths
$\lambda k_\mathrm{F}/E_\mathrm{F} > 1$, we still find signatures of a
Fermi surface persisting throughout the TSF phase, and canonical
BCS-like behavior being exhibited for $h\gtrsim h_\mathrm{max}$.

The remainder of this article is organized as follows.
Section~\ref{sec:formalism} introduces the theoretical approach used
by us to describe the BCS-to-BEC crossover for the
\textit{s}-wave-paired 2D Fermi gas subject to both spin-orbit
coupling and Zeeman spin splitting. Detailed results obtained within
this formalism for the system with fixed uniform particle density are
presented in the subsequent Sec.~\ref{sec:results}, together with a
discussion of physical implications and limitations inherent in the
mean-field approach. Our conclusions are formulated in the final
Sec.~\ref{sec:concl}.

\section{Microscopic model of the 2D TSF}\label{sec:formalism}

We utilize a standard Bogoliubov-de Gennes (BdG) mean-field
formalism~\cite{deGennes1989} to calculate the quasiparticle spectrum
for our system of interest. All relevant thermodynamic quantities can
be expressed in terms of the obtained eigenenergies and eigenstates.
Throughout this work, we consider the zero-temperature limit.

The  BdG Hamiltonian of the 2D spin-orbit coupled Fermi gas with
$s$-wave interactions and Zeeman spin splitting $2h$ acting in the
four-dimensional Nambu space of spin-$1/2$ fermions is~\footnote{Our
notation adheres to that used in Ref.~\cite{Brand2018}.}
\begin{align}\label{eq:origBdG}
\mathcal{H} = \left( \begin{array}{cccc} \epsilon_{\kk \uparrow}
-\mu &\lambda_\kk & 0& -\Delta \\ {\lambda_\kk}^* & \epsilon_{\kk
\downarrow} - \mu &\Delta & 0 \\ 0 & \Delta^* & -\epsilon_{\kk
\uparrow} + \mu & {\lambda_\kk^*} \\ -\Delta^* & 0 &\lambda_\kk &
-\epsilon_{\kk\downarrow} +\mu \end{array} \right) ,
\end{align}
where $\kk = (k_x, k_y)$ denotes the 2D wave vector, $\epsilon_{\kk
\uparrow (\downarrow)}=\epsilon_\kk \, \substack{-\\(+)} \, h$ with
$\epsilon_\kk = \hbar^2 (k_x^2+k_y^2)/2 m$, and $\lambda_\kk \equiv
\lambda\, i(k_x - i k_y)$ is the spin-orbit coupling~\footnote{While
we adopt the 2D-Rashba form~\cite{Bychkov1984} for $\lambda_\kk $, our
results apply also to other types of spin-orbit coupling that depend
linearly on the components of $\kk$, such as the 2D-Dirac and
2D-Dresselhaus functional forms~\cite{Winkler2003,Galitski2013}
corresponding to $\lambda_\kk =\lambda (k_x-i k_y)$ and $\lambda (k_x
+ i k_y)$, respectively.}. The BdG equation reads
\begin{align}\label{eq:bdg}
\mathcal{H} \left(\begin{array}{c} u^\uparrow \\ u^\downarrow \\
v^\uparrow \\ v^\downarrow \end{array}\right)  = E \left(
\begin{array}{c} u^\uparrow \\ u^\downarrow \\ v^\uparrow \\
v^\downarrow \end{array}\right) \quad .
\end{align}
Its spectrum consists of four eigenvalue  branches~\cite{Yi2011,
Zhou2011},
\begin{widetext}
\begin{equation}\label{eq:fullSpec}
E_{\kk\alpha,<(>)} = \alpha \sqrt{(\epsilon_\kk -\mu)^2 +|\Delta|^2 +
h^2 + |\lambda_\kk|^2 \substack{-\\(+)} 2 \sqrt{(\epsilon_\kk -\mu)^2
( h^2 + |\lambda_\kk|^2) + |\Delta|^2 h^2}} \quad ,
\end{equation}
\end{widetext}
with associated eigenspinors $(u^\uparrow_{\kk\alpha,\eta},
u^\downarrow_{\kk\alpha,\eta}, v^\uparrow_{\kk\alpha,\eta},
v^\downarrow_{\kk\alpha,\eta})^T$, where $\alpha \in \{+,-\}$ and
$\eta \in \{<,>\}$ label the four different energy-dispersion
branches.

The chemical potential $\mu$ and magnitude $|\Delta|$ of the pair
potential need to be determined selfconsistently from solutions of
the BdG equations in conjunction with the gap equation and the
constraint that the uniform particle density is fixed at $n\equiv
k_\mathrm{F}^2/(2\pi)$. Corresponding conditions can be formulated
mathematically in terms of the energy spectrum and BdG-Hamiltonian
eigenspinor amplitudes. See, e.g., Refs.~\cite{deGennes1989,
Brand2018}. However, educated by the insights gained from previous
work on spin-imbalanced Fermi superfluids~\cite{Radzihovsky2010}, we
base selfconsistency considerations on the properties of the system's
grand-canonical ground-state energy density~\cite{Sheehy2007,
Parish2007,Radzihovsky2010,Yi2011,Zhou2011}, for which a standard
calculation yields
\begin{eqnarray}\label{eq:gsE}
&& \mathcal{E}_\mathrm{gs}^\mathrm{(MF)}(|\Delta|, \mu) = \nonumber
\\ && \hspace{1cm} \frac{1}{A}\sum_\kk \Big( \frac{|\Delta|^2}{2
\epsilon_\kk + E_\mathrm{b}} +\epsilon_\kk -\mu -\frac{1}{2} \sum_\eta
E_{\kk+,\eta} \Big)\, . \quad
\end{eqnarray}
Here $A$ denotes the system's volume (area), and $E_\mathrm{b}>0$ is
the magnitude of the two-particle bound-state (i.e., binding) energy
\cite{noteEb}. The gap and number-density equations can be expressed
in terms of derivatives of the ground-state energy density;
\begin{subequations}
\begin{eqnarray}\label{eq:minEgs}
\frac{\partial \mathcal{E}_\mathrm{gs}^\mathrm{(MF)}}{\partial
|\Delta|} &=& 0 \quad , \\[0.1cm]
\frac{\partial \mathcal{E}_\mathrm{gs}^\mathrm{(MF)}}{\partial\mu}
&=& - n \quad . \label{eq:denEgs}
\end{eqnarray}
\end{subequations}
The lengthy explicit expressions are omitted here.

As emphasized previously during the study of spin-imbalanced Fermi
superfluids~\cite{Sheehy2007a}, proper application of the condition
(\ref{eq:minEgs}) for identifying physical ground states requires
ensuring that $\mathcal{E}_\mathrm{gs}^\mathrm{(MF)}(|\Delta|, \mu)$,
taken as a function of $|\Delta|$ at fixed $\mu$, has a global minimum
at the selfconsistently determined value for $|\Delta|$. However,
identifying local minima as well as maxima of the ground-state energy
at fixed $\mu$ can also be of interest~\cite{Sarma1963,Lamacraft2008,
He2009}, e.g., to discuss nonequilibrium-dynamic phenomena; hence, we
will track these in the following also.

The relative magnitude of $E_\mathrm{b}$ with respect to the Fermi
energy $E_\mathrm{F}$ drives the BCS-to-BEC crossover for
\textit{s}-wave pairing in our system of interest~\cite{Randeria1990}.
More specifically, we have
\begin{equation}
\frac{E_\mathrm{b}}{E_\mathrm{F}} \left\{ \begin{array}{cl} \ll 1 &
\mbox{in the BCS limit,} \\[0.2cm] \gtrsim 1 & \mbox{in the BEC
regime.} \end{array} \right.
\end{equation}
In the following, we absorb any dependence on total particle density
$n$ by measuring all energies and wave vectors in units of
$E_\mathrm{F}$ and $k_\mathrm{F}$, respectively. Thus the set of
externally tuneable parameters comprises $E_\mathrm{b}/E_\mathrm{F}$,
$h/E_\mathrm{F}$, and $\lambda k_\mathrm{F}/E_\mathrm{F}$. The
system's state is characterized by $|\Delta|/E_\mathrm{F}$ and
$\mu/E_\mathrm{F}$.

The chiral $p$-wave nature of the superfluid is revealed by the
following considerations. Inspection of Eq.~(\ref{eq:fullSpec}) shows
that $E_{\vek{0} +,<} = | h_\mathrm{c} - h |$. In the BCS regime, for
$0< h < h_\mathrm{c}$, two minima exist in $E_{\kk +,<}$ at $|\kk| >
0$, corresponding to effective \textit{p}-wave pairing around the two
spin-split Fermi surfaces for spin-$\uparrow$ and spin-$\downarrow$
degrees of freedom. As $h$ is increased, the location of the
spin-$\downarrow$ minimum moves towards $|\kk|=0$, with its value
shrinking and finally vanishing as it reaches $|\kk|=0$ at $h =
h_\mathrm{c}$. For $h > h_\mathrm{c}$, the system has only one Fermi
surface corresponding to a fully polarized electron system, and the
remaining minimum of $E_{\kk +,<}$ at $|\kk| \sim \sqrt{2}
k_\mathrm{F}$ is associated with an effective pair
potential~\cite{Sato2010,Brand2018} $(\lambda_\kk/|\lambda_\kk|) \,
\Delta_\mathrm{pw}\equiv i \ee^{- i\varphi_\kk}\, \Delta_\mathrm{pw}$,
where $\varphi_\kk =\arctan(k_y/k_x)$ is the polar-angle coordinate
for the 2D wave vector $\kk$. Proportionality of the superconducting
order parameter to the phase factor $\ee^{-i \varphi_\kk}$ is the
defining property of chiral \textit{p}-wave pairing~\cite{Kallin2016},
and also the origin of its accompanying topological
features~\cite{Kallin2016,Sato2017}. In contrast, the system has
\emph{two\/} Fermi surfaces where \textit{p}-wave pairing with
\emph{opposite\/} chirality occurs when $h < h_\mathrm{c}$, rendering
it to be a nontopological superfluid. We now apply the formalism
introduced above to investigate the fate of chiral \textit{p}-wave
superfluidity in the BEC regime for the underlying \textit{s}-wave
pairing.

\section{Results and discussion}\label{sec:results}

\begin{figure}[t]
\includegraphics[width=1.0\columnwidth]{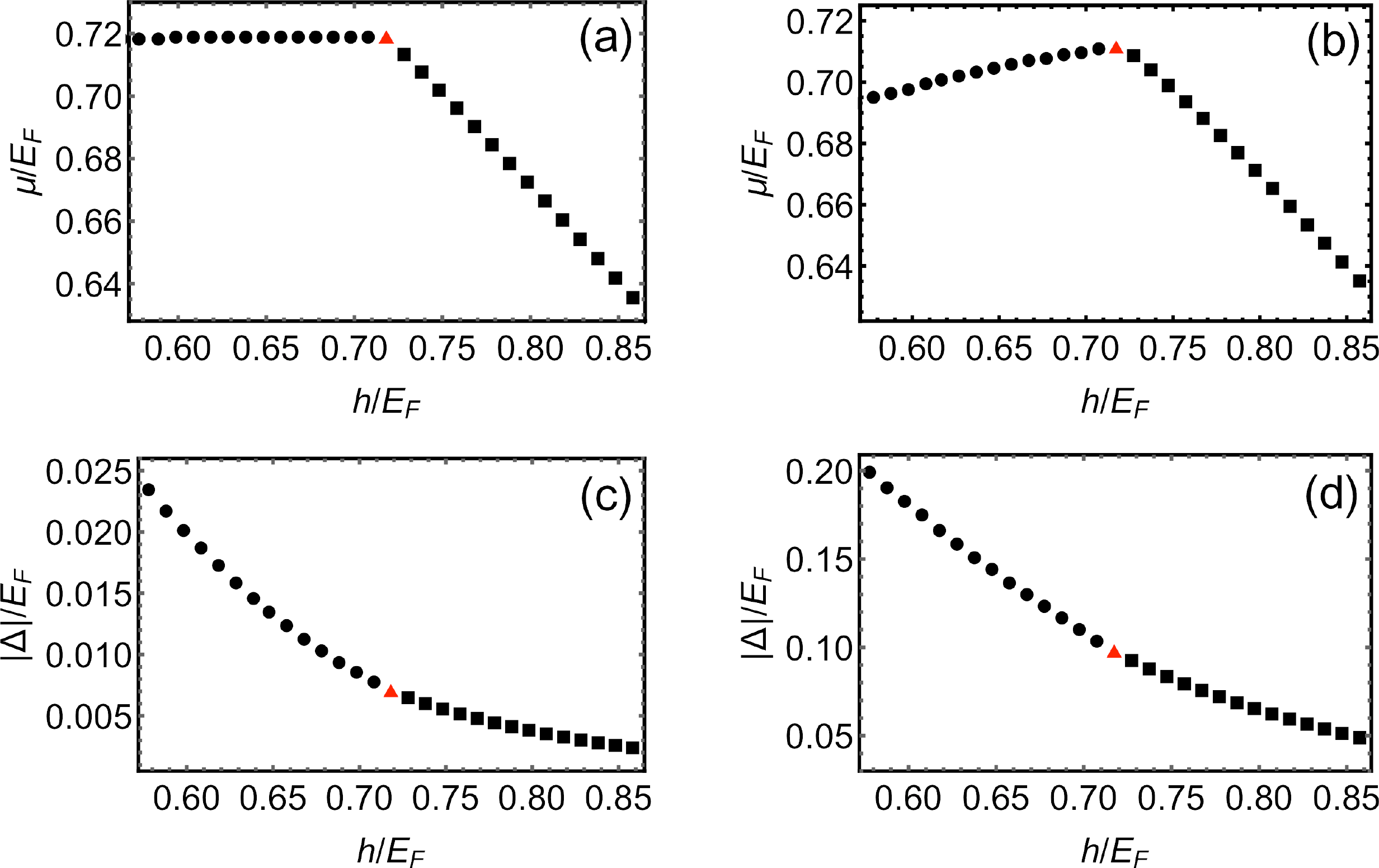}
\caption{\label{fig:BCSlim}%
Chemical potential $\mu$ and magnitude $|\Delta|$ of the
\textit{s}-wave pair potential for a spin-orbit-coupled 2D Fermi
system with fixed density $n = m E_\mathrm{F}/(\pi\hbar^2)\equiv
k_\mathrm{F}^2/(2\pi)$ in the BCS regime for \textit{s}-wave pairing,
plotted as a function of Zeeman splitting $h$. Results shown are
obtained as solutions of the selfconsistency conditions
[Eqs.~(\ref{eq:minEgs}) and (\ref{eq:denEgs})] for $\lambda
k_\mathrm{F}/E_\mathrm{F} = 0.75$ (all panels) and $E_\mathrm{b}/
E_\mathrm{F} = 0.010$ [panels (a) and (c)], $0.10$ [panels (b) and
(d)]. Data points indicated by circles (a triangle, squares)
correspond to states where the system is nontopological (critical,
topological), i.e., $h < (=, >)\,\sqrt{\mu^2 + |\Delta|^2}$.}
\end{figure}

\begin{figure}[t]
\includegraphics[width=1.0\columnwidth]{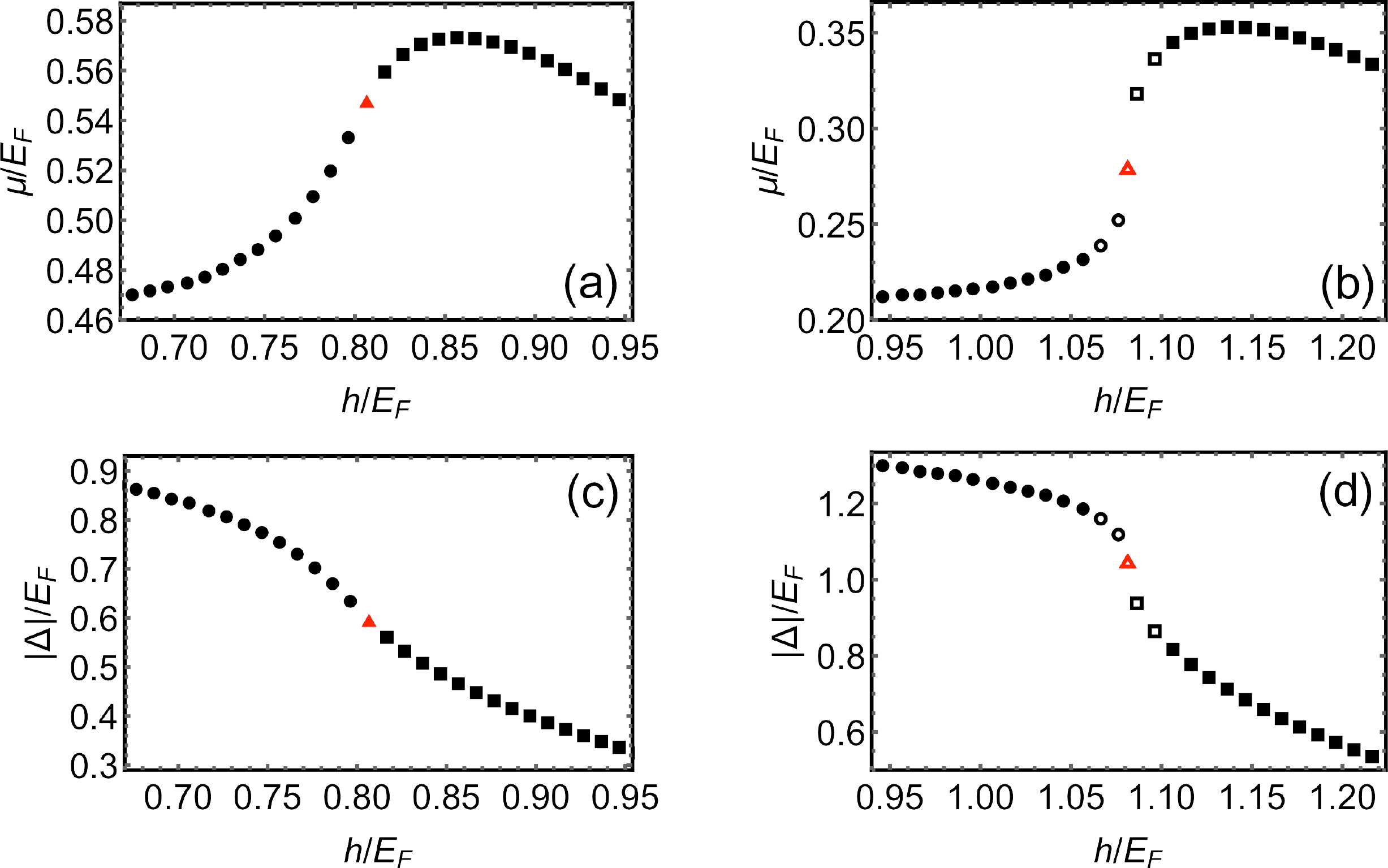}
\caption{\label{fig:BECclose1}%
Chemical potential $\mu$ and \textit{s}-wave pair-potential magnitude
$|\Delta|$ for a spin-orbit-coupled 2D Fermi system with fixed density
approaching the BEC regime for \textit{s}-wave pairing. Results shown
are obtained as solutions of the selfconsistency conditions for
$\lambda k_\mathrm{F}/E_\mathrm{F} = 0.75$ (all panels) and
$E_\mathrm{b}/E_\mathrm{F}=0.50$ [panels (a) and (c)], $1.0$ [panels
(b) and (d)]. Circles (a triangle, squares) correspond to states where
the system is nontopological (critical, topological). Filled (empty)
symbols indicate solutions of the selfconsistency conditions that
globally minimize the ground-state energy and thus correspond to
proper equilibrium states of the system (that represent only a local
minimum or even a maximum of the ground-state energy).}
\end{figure}

\begin{figure}[b]
\includegraphics[width=0.85\columnwidth]{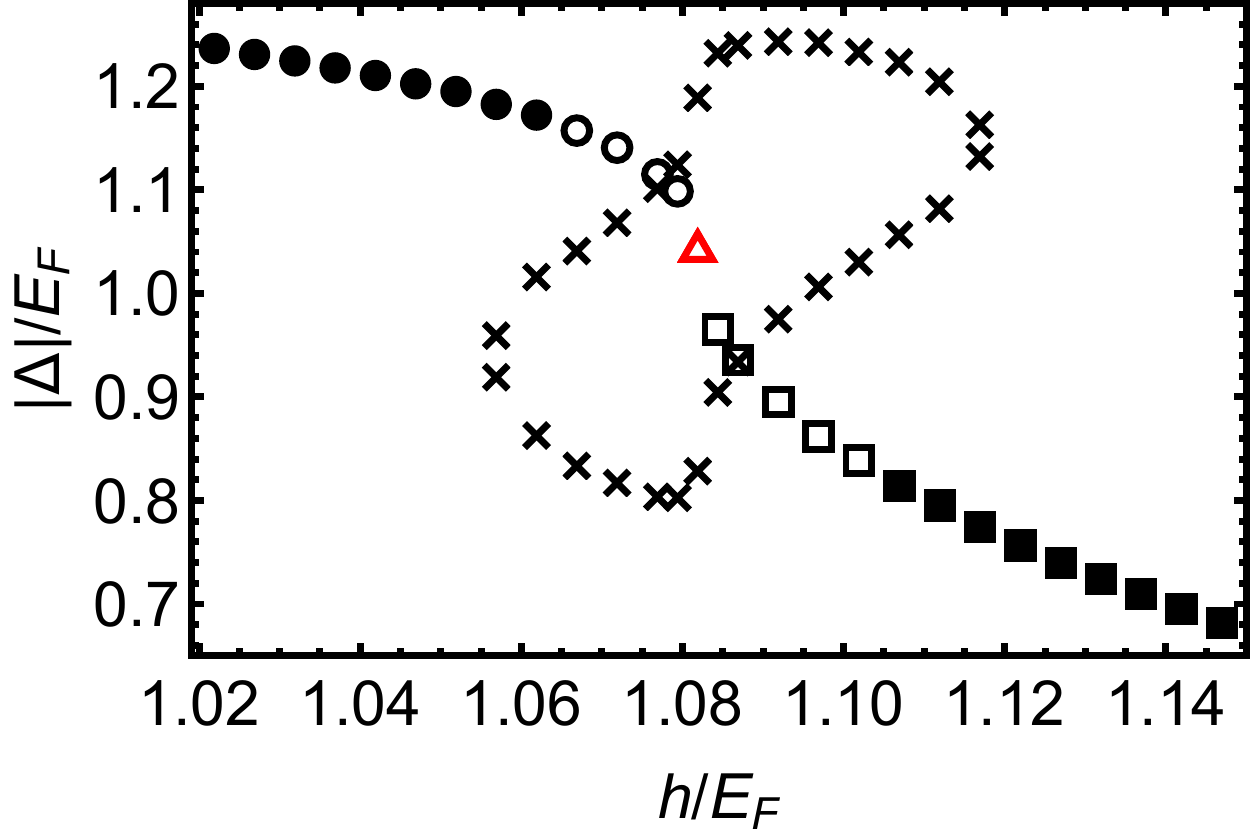}
\caption{\label{fig:BECclose2}%
Emergence of additional zeros of the gap equation in the BEC regime.
Results shown are obtained for $\lambda k_\mathrm{F}/E_\mathrm{F} =
0.75$ and $E_\mathrm{b}/E_\mathrm{F} = 1.0$. Filled symbols indicate
solutions of the selfconsistency conditions that globally minimize
the ground-state energy and thus correspond to proper equilibrium
states of the system. Empty symbols (crosses) are associated with
(non)selfconsistent values corresponding to a local minimum or
maximum of the ground-state energy. Circles (a triangle, squares)
indicate selfconsistent solutions where the system is nontopological
(critical, topological).}
\end{figure}

\begin{figure*}[t]
\includegraphics[width=1.0\textwidth]{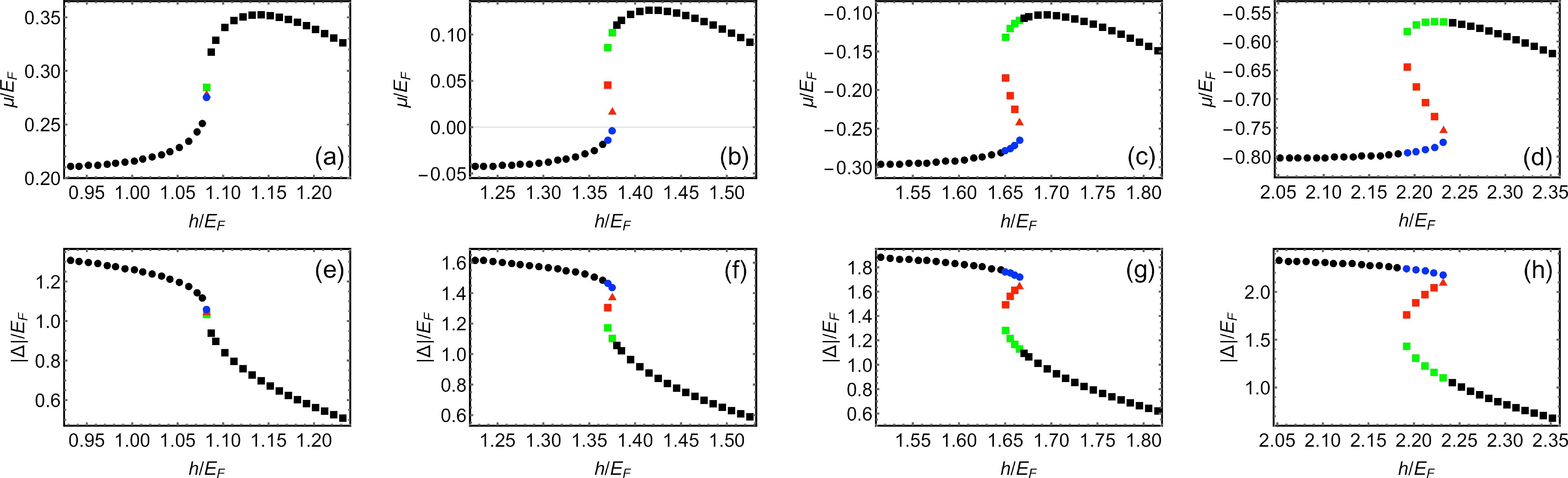}
\caption{\label{fig:multSC}%
Emergence of multiple pairs of selfconsistent solutions for the
chemical potential $\mu$ and \textit{s}-wave pair-potential magnitude
$|\Delta|$, indicated by colors. Results shown here are obtained for
$\lambda k_\mathrm{F}/E_\mathrm{F} = 0.75$ (all panels) and
$E_\mathrm{b}/E_\mathrm{F}=1.001$ [panels (a) and (e)], $1.5$ [panels
(b) and (f)], $2.0$ [panels (c) and (g)], $3.0$ [panels (d) and (h)].
Circles (a triangle, squares) correspond to states where the system is
nontopological (critical, topological).}
\end{figure*}

To ground ourselves in well-known results~\cite{He2013,Brand2018}, we
start by fixing a value for $\lambda k_\mathrm{F}/E_\mathrm{F}$ and
consider the variation of the chemical potential $\mu$ and the
pair-potential magnitude $|\Delta|$ as a function of the Zeeman energy
$h$ in the BCS limit for \textit{s}-wave pairing, i.e., for small
$E_\mathrm{b}/E_\mathrm{F}$. As illustrated in Fig.~\ref{fig:BCSlim},
both $\mu(h)$ and $|\Delta(h)|$ evolve continuously from the
nontopological phase where $h < h_\mathrm{c}$ [defined in
Eq.~(\ref{eq:critZeem})] via their critical values $\mu_\mathrm{c}
\equiv \mu(h_\mathrm{c})$ and $|\Delta(h_\mathrm{c})| \equiv
\Delta_\mathrm{c}$ that satisfy $\sqrt{\mu_\mathrm{c}^2 +
\Delta_\mathrm{c}^2} = h_\mathrm{c}$ into the topological phase where
$h > h_\mathrm{c}$. This reflects the fact that, for any value of $h$,
${\mathcal E}_\mathrm{gs}^\mathrm{(MF)}$ has only a single minimum
when plotted as a function of $|\Delta|$ for fixed $\mu$, which occurs
at a nonzero $|\Delta|$ and thus corresponds to a homogeneous
superfluid ground state.

\begin{figure}[b]
\includegraphics[width=0.85\columnwidth]{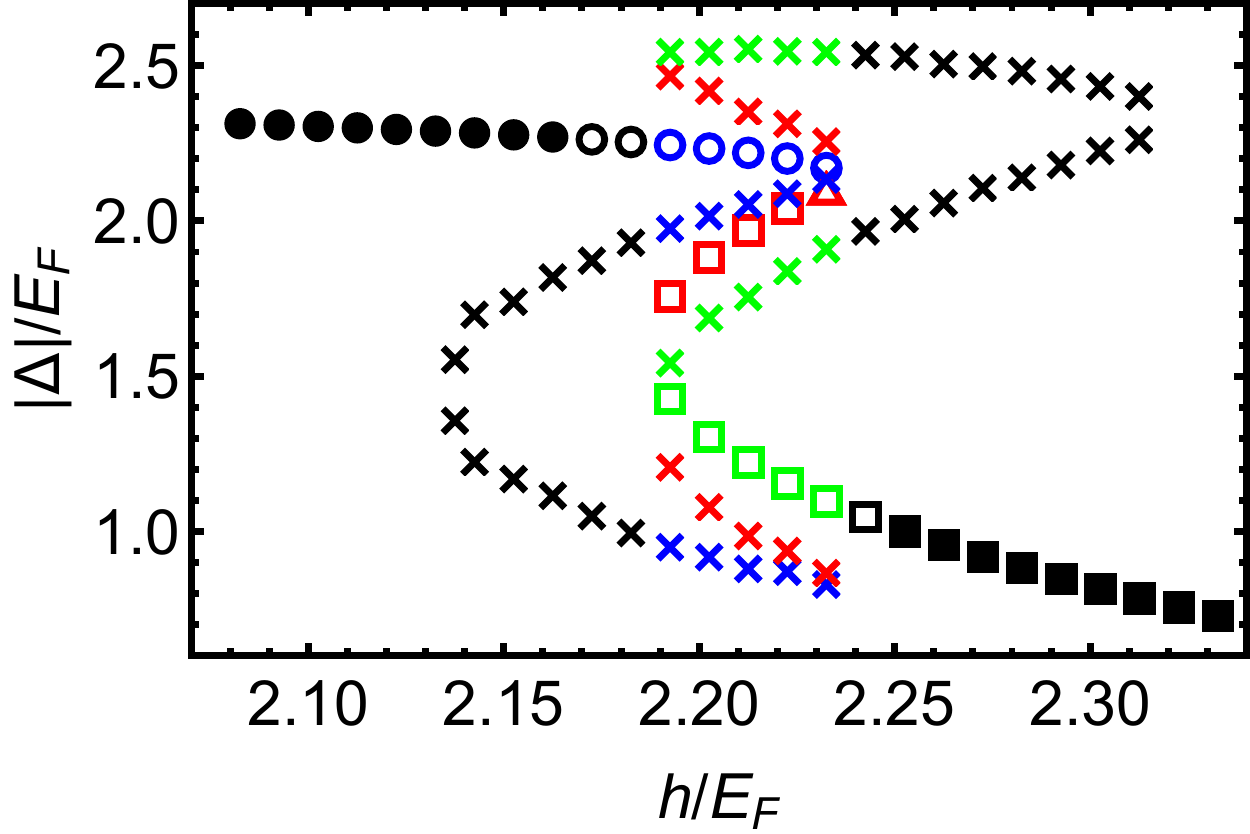}
\caption{\label{fig:BEClim}%
Structure of multiple selfconsistent and associated
nonselfconsistent solutions of the gap equation deep in the BEC
regime. Results shown are obtained for $\lambda k_\mathrm{F}/
E_\mathrm{F} = 0.75$ and $E_\mathrm{b}/E_\mathrm{F} = 3.0$. Filled
symbols indicate solutions of the selfconsistency conditions that
globally minimize the ground-state energy and thus correspond to
proper equilibrium states of the system. Empty symbols (crosses) are
associated with (non)selfconsistent values corresponding to a local
minimum or maximum of the ground-state energy. Circles (a triangle,
squares) indicate states where the system is nontopological (critical,
topological). Multiple selfconsistent solutions at a given $h$ are
distinguished by color. The same color is used to indicate their
associated additional zeros in the gap equation.}
\end{figure}

The search for solutions of the selfconsistency conditions
(\ref{eq:minEgs}) and (\ref{eq:denEgs}) for larger $E_\mathrm{b}/
E_\mathrm{F}\lesssim 1$ continues to yield unique values of $|\Delta|$
and $\mu$. See the examples shown in Fig.~\ref{fig:BECclose1}.
However, an intricate complexity associated with selfconsistent
solutions starts to develop. As illustrated in
Fig.~\ref{fig:BECclose2}, within an intermediate range of Zeeman
energies, two additional extrema (specifically, a local minimum and a
local maximum) start to appear in the $|\Delta|$-dependence of the
ground-state energy where $\mu$ has been fixed to its selfconsistent
value. Below the value $E_\mathrm{b}^{(\mathrm{c})}$ associated with
the critical end-point of the phase-separation region shown in
Fig.~\ref{fig:phaseDia}, the unique solution of the selfconsistency
conditions still continues to be the global minimum of ${\mathcal
E}_\mathrm{gs}^\mathrm{(MF)}$, taken at the selfconsistent $\mu$, for
any value of $h$. This is the case, e.g., for the system parameters
used to calculate the results shown in Fig.~\ref{fig:BECclose1}(a,c).
However, for $E_\mathrm{b}\ge E_\mathrm{b}^{(\mathrm{c})}$, which
applies to Fig.~\ref{fig:BECclose1}(b,d), the selfconsistently
determined value for $|\Delta|$ ceases to be associated with the
global minimum of ${\mathcal E}_\mathrm{gs}^\mathrm{(MF)}$ at fixed
selfconsistent $\mu$ for Zeeman energies within a range $h_< < h <
h_>$, corresponding instead to only a local minimum or even a
maximum. This implies that no single-phase equilibrium ground state
exists in the region $h_< < h < h_>$. Instead,  phase separation into
domains of different densities will occur if the system is driven into
this region. Even further in the BEC regime when $E_\mathrm{b}>
E_\mathrm{b}^{(\mathrm{m})}$, multiple selfconsistent pairs of values
for $|\Delta|$ and $\mu$ emerge as illustrated in
Fig.~\ref{fig:multSC}. Around each of these, additional zeros of the
gap equation exist, as seen in Fig.~\ref{fig:BEClim}. Now the range
$h_< < h < h_>$ is defined to be the region where none of the
selfconsistent $|\Delta|$ values is associated with the global
minimum of the ground-state energy ${\mathcal
E}_\mathrm{gs}^\mathrm{(MF)}$ when $\mu$ is fixed to its corresponding
selfconsistent value.

\begin{figure}[t]
\includegraphics[height=3.1cm]{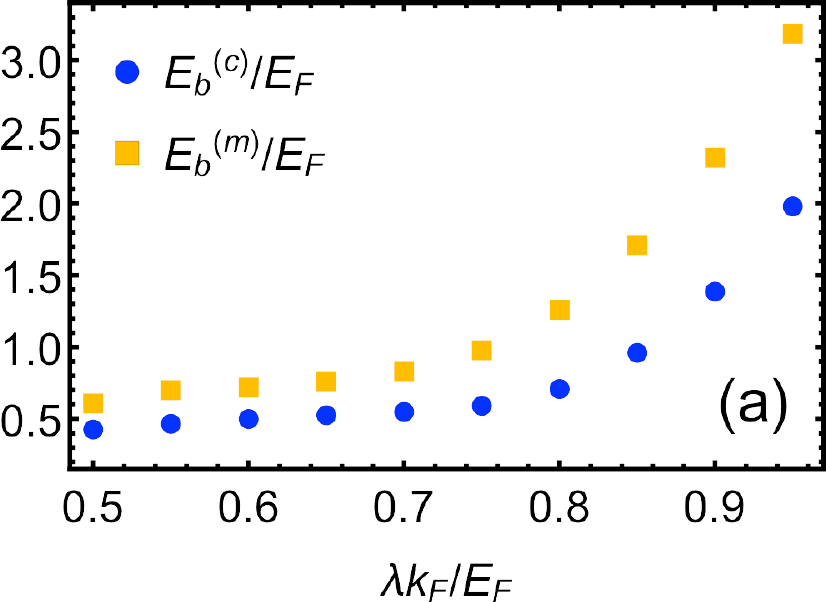}\hfill
\includegraphics[height=3.1cm]{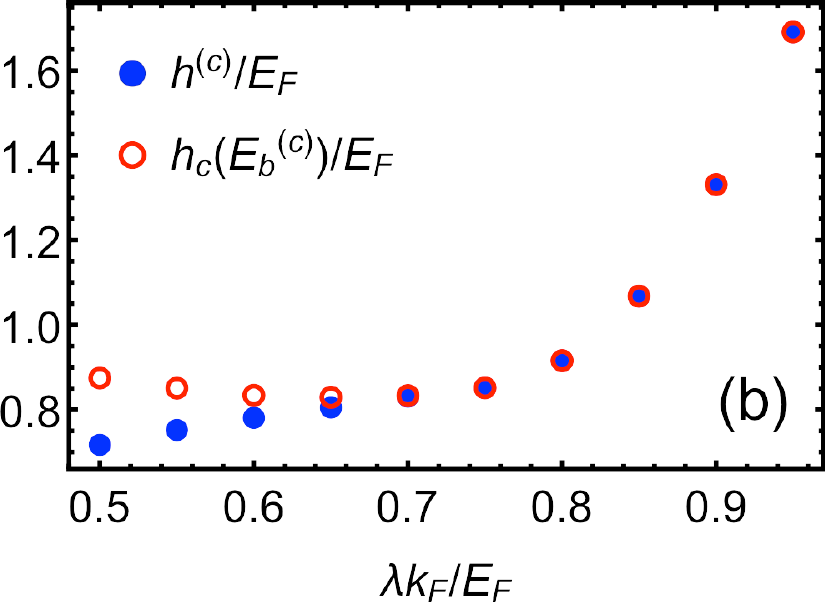}
\caption{\label{fig:critEbS}%
Panel (a): Special values $E_\mathrm{b}^{(\mathrm{c})}$ and
$E_\mathrm{b}^{(\mathrm{m})}$ for the two-particle bound-state energy
$E_\mathrm{b}$ plotted as a function of the spin-orbit-coupling
strength $\lambda$. Here $E_\mathrm{b}^{(\mathrm{c})}$ is the value
of $E_\mathrm{b}$ associated with the critical end point $\big(
h^\mathrm{(c)}, E_\mathrm{b}^{(\mathrm{c})} \big)$ of the
phase-separation region in the $E_\mathrm{b}$-$h$ phase diagram where
the critical-field curves $h_<$ and $h_>$ merge. The value
$E_\mathrm{b}^{(\mathrm{m})}$ is the lower limit of bound-state
energies for which multiple pairs of selfconsistent solutions for
$\mu$ and $|\Delta|$ exist. Panel (b): Dependence of $h^\mathrm{(c)}$,
the $h$ coordinate of the critical end point of the phase-separation
region in the $E_\mathrm{b}$-$h$ phase diagram, on the
spin-orbit-coupling strength. For comparison, the critical field
$h_\mathrm{c}$ [defined in Eq.~(\ref{eq:critZeem})] at
$E_\mathrm{b}^{(\mathrm{c})}$ is also shown.}
\end{figure}

The appearance of multiple extrema in the $|\Delta|$ dependence of
${\mathcal E}_\mathrm{gs}^\mathrm{(MF)}$ at fixed $\mu$, leading to
the selfconsistent minimum ceasing to be the global minimum,
indicates the presence of a first-order (noncontinuous) phase
transition~\cite{He2008,Zhou2011}. A proper theoretical description of
this situation requires the construction of various phase-coexistence
scenarios~\cite{Yi2011,Yang2012, Seo2012}, in analogy with treatments
developed for the population-imbalanced Fermi gas without spin-orbit
coupling~\cite{Bedaque2003,Carlson2005,Sheehy2006,Son2006,Sheehy2007,
Parish2007,He2008,Du2009}. Here we defer the careful determination of
the equilibrium ground state in the phase-separation region to future
work~\footnote{This task becomes particularly challenging for the part
of the phase diagram where multiple selfconsistent solutions of the
gap equation exist at fixed $h$. Generally, two of these correspond to
minima of the ground-state energy taken at fixed $\mu$, and their
combined evolution between global- or local-minimum status needs to be
tracked.}. Rather, we intend to discuss the properties of the adjacent
uniform, single-phase regions for large $E_\mathrm{b}/E_\mathrm{F}$.
To this end, we only need to map carefully the boundaries of the
phase-separation region, i.e., the critical-Zeeman-energy curves $h_<$
and $h_>$. Results for representative values of the
spin-orbit-coupling strength are given in Fig.~\ref{fig:phaseDia}. We
find that the phase-separation region narrows as the
spin-orbit-coupling parameter $\lambda k_\mathrm{F}/E_\mathrm{F}$ is
increased, while simultaneously the critical end point $\big(
h^{(\mathrm{c})}, E_\mathrm{b}^{(\mathrm{c})} \big)$ where the $h_<$
and $h_>$ curves merge shifts to larger coordinate values in the phase
diagram. The full dependence of $E_\mathrm{b}^{(\mathrm{c})}$ (and
also of $E_\mathrm{b}^{(\mathrm{m})}$) as a function of the
dimensionless spin-orbit-coupling strength is plotted in
Fig.~\ref{fig:critEbS}(a), with the associated results for
$h^{(\mathrm{c})}$ being provided in Fig.~\ref{fig:critEbS}(b). Two
different regimes, corresponding to small and large values of $\lambda
k_\mathrm{F}/E_\mathrm{F}$, can be identified, where the former
(latter) is characterized by the $h^{(\mathrm{c})}$ values diverging
from (coinciding with) the critical field $h_\mathrm{c}$ for
$E_\mathrm{b}=E_\mathrm{b}^{(\mathrm{c})}$.

The curves for $h_<(E_\mathrm{b})$ and $h_>(E_\mathrm{b})$ in the
phase diagram delimit the phase-separation region associated with a
first-order transition between different superfluid states. In those
parts of the phase diagram outside this region where only a single
pair of selfconsistent values for $\mu$ and $|\Delta|$ exists, a
curve $h_\mathrm{c}(E_\mathrm{b})$ can be defined via
Eq.~(\ref{eq:critZeem}) that separates the part of the phase diagram
where the system is an ordinary nontopological superfluid (NSF, for
$h<h_\mathrm{c}$) from the part where the ground state corresponds to
a topological superfluid (TSF, for $h>h_\mathrm{c}$). In particular,
for $E_\mathrm{b}<E_\mathrm{b}^{(\mathrm{c})}$, only this second-order
topological transition occurs. However, beyond the point where the
curve for $h_\mathrm{c}(E_\mathrm{b})$ crosses that of $h_>
(E_\mathrm{b})$, solutions of the selfconsistency conditions that are
critical, i.e., satisfy $h=\sqrt{\mu^2 + |\Delta|^2}$, continue to
exist but are no longer a global minimum of the ground-state energy at
fixed $\mu$. At the same time, the homogeneous-superfluid states
existing for $h>h_>$ satisfy $h>\sqrt{\mu^2 + |\Delta|^2}$ and are
thus in the topological phase. Hence, beyond the crossing point of
$h_\mathrm{c}(E_\mathrm{b})$ and $h_>(E_\mathrm{b})$, the topological
transition is of first order. The phase boundary of the homogeneous
2D TSF is therefore delineated by $h_{\text{max}} (E_\mathrm{b})=
\mathrm{max}\{h_\mathrm{c}(E_\mathrm{b}), h_>(E_\mathrm{b})\}$. Due
to the tendency of $|\Delta|$ to monotonically decrease with $h$ in
regions where a selfconsistent solution is associated with the
system's equilibrium ground state (see Figs.~\ref{fig:BCSlim},
\ref{fig:BECclose2}, and \ref{fig:BEClim}), $h_\mathrm{max}$ is also
the Zeeman energy for which $|\Delta|$ is maximized in the TSF phase
at fixed $E_\mathrm{b}$. We now focus on the properties of the
single-phase ground states adjacent to the phase-separation region
at large $E_\mathrm{b} > E_\mathrm{b}^{(\mathrm{c})}$.

\begin{figure}[t]
\includegraphics[width=1.0\columnwidth]{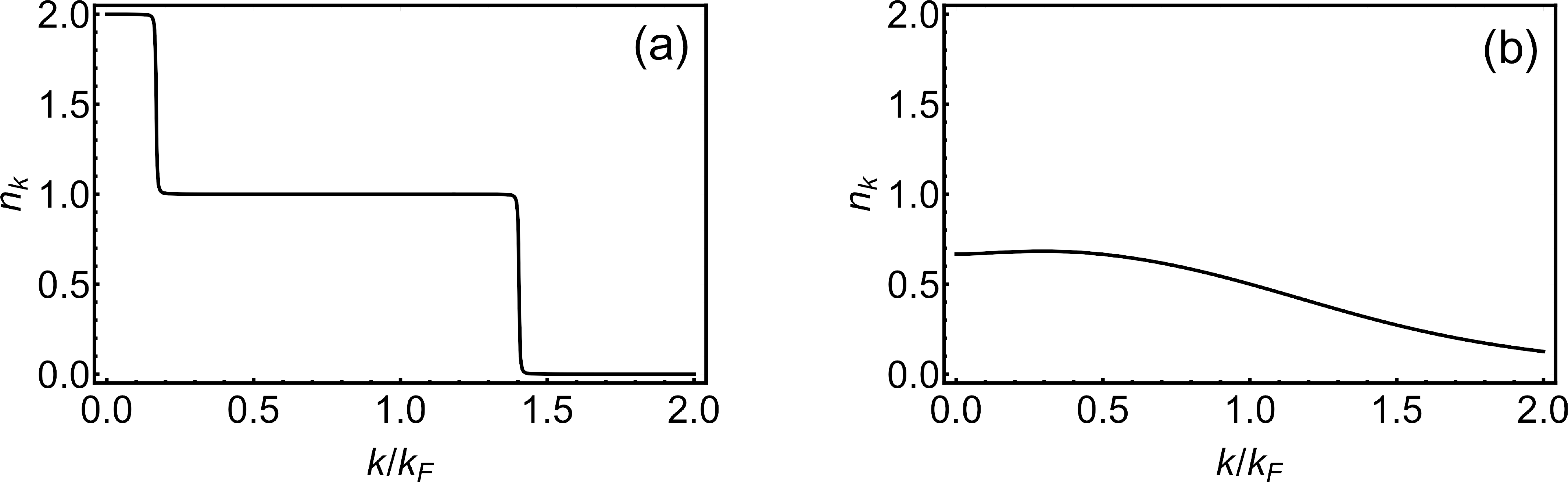}\\[0.2cm]
\hfill\includegraphics[width=0.98\columnwidth]{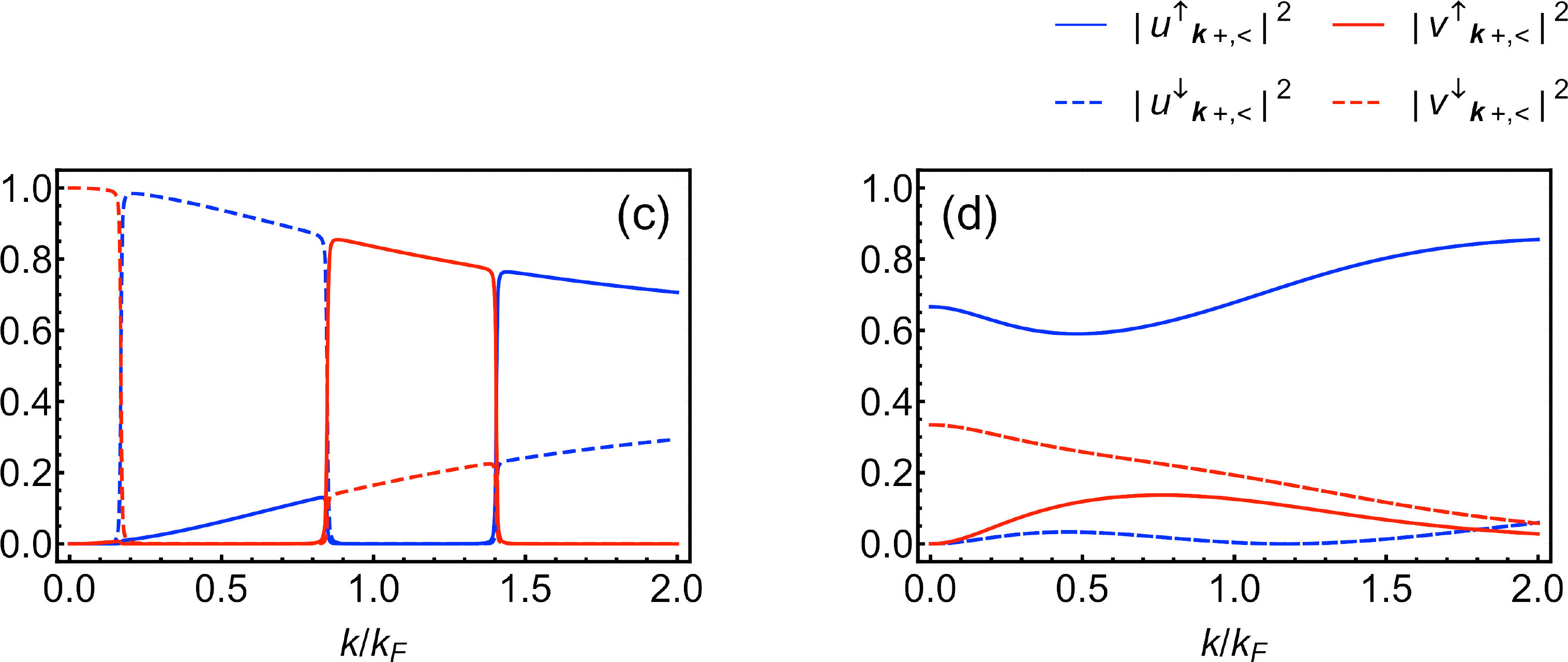}\\[0.2cm]
\hfill\includegraphics[width=0.99\columnwidth]{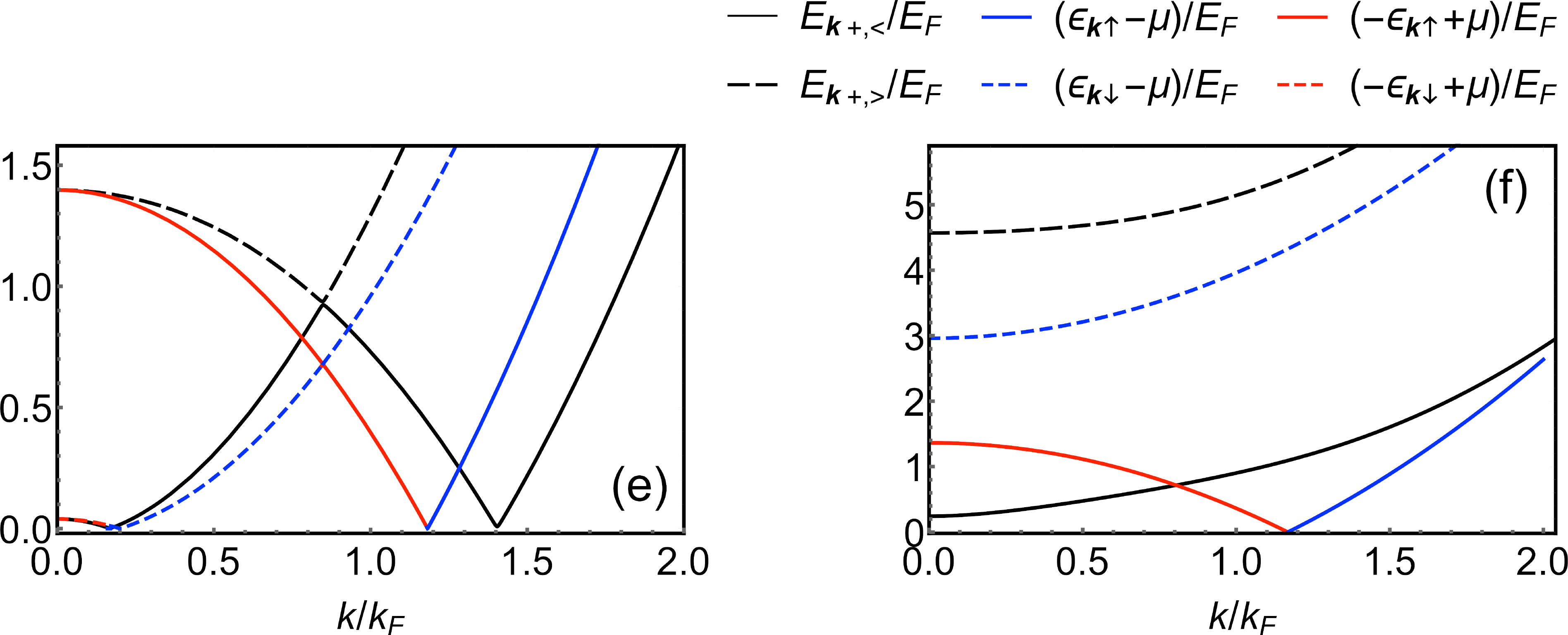}

\vspace{-0.1cm}
\caption{\label{fig:trivial}%
Expected BCS and BEC characteristics are exhibited in the
nontopological-superfluid (NSF) phase of a spin-orbit-coupled 2D Fermi
gas. All results plotted here have been obtained for fixed $\lambda
k_\mathrm{F}/E_\mathrm{F} = 0.75$, and $k\equiv |\kk|$ is the
magnitude of the 2D wave vector of Bogoliubov quasiparticles. Panels
(a), (c) and (e) depict the BCS regime, showing results for
$E_\mathrm{b}/E_\mathrm{F}=0.010$ and $h = h_\mathrm{c} - 0.04\,
E_\mathrm{F}$. Juxtaposed are panels (b), (d) and (f) associated with
the BEC regime, specifically for $E_\mathrm{b}/E_\mathrm{F}=3.0$ and
$h = h_<$ (corresponding to the filled circle with maximum $h$ in
Fig.~\ref{fig:BEClim}). The presence [absence] of Fermi-surface
features in the momentum-space density distribution $n_k$ displayed in
panel (a) [(b)] is typical for the BCS [BEC] regime of \textit{s}-wave
pairing.  Also the $k$ dependence of spin-resolved Bogliubov
amplitudes shown in panels (c) and (d) exhibits the familiar pattern,
with the spin-$\uparrow$ Nambu-particle component becoming dominant in
the BEC regime. The quasiparticle dispersion $E_{\kk +, <}$ in the BCS
regime [shown in panel (e)] has two minima corresponding to excitation
gaps at finite $k$ but, as expected, the excitation gap is at $k=0$ in
the BEC regime [see panel (f)]. Abrupt changes in the spin-resolved
Bogoliubov amplitudes in the BCS regime [panel (c)] are the result of
small-gap anticrossings between highly spin-polarized dispersion
branches. We illustrate this by showing also the purely Zeeman-split
dispersions $|\epsilon_{\kk\sigma}-\mu|$, which quite closely resemble
the full dispersions $E_{\kk +, \lessgtr}$ in the BCS regime [panel
(e)]. In contrast, $E_{\kk +, <}$ is qualitatively different from
$|\epsilon_{\kk\uparrow}-\mu|$ in the BEC regime [panel (f)].}
\end{figure}

\begin{figure}[t]
\includegraphics[width=1.0\columnwidth]{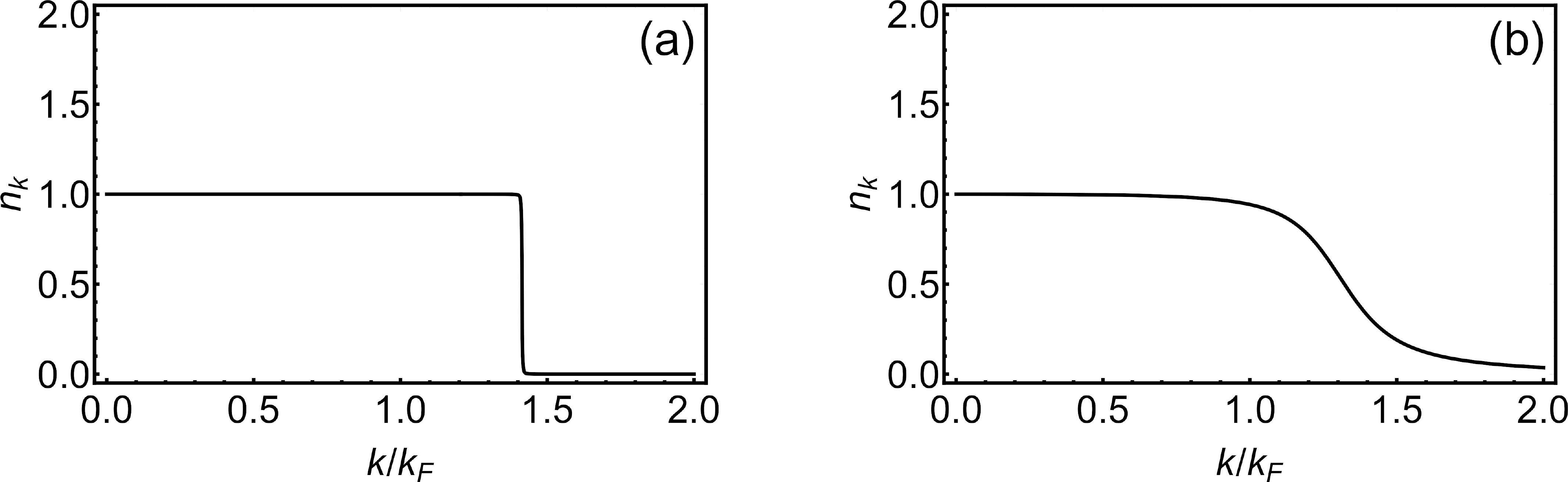}\\[0.2cm]
\hfill\includegraphics[width=0.97\columnwidth]{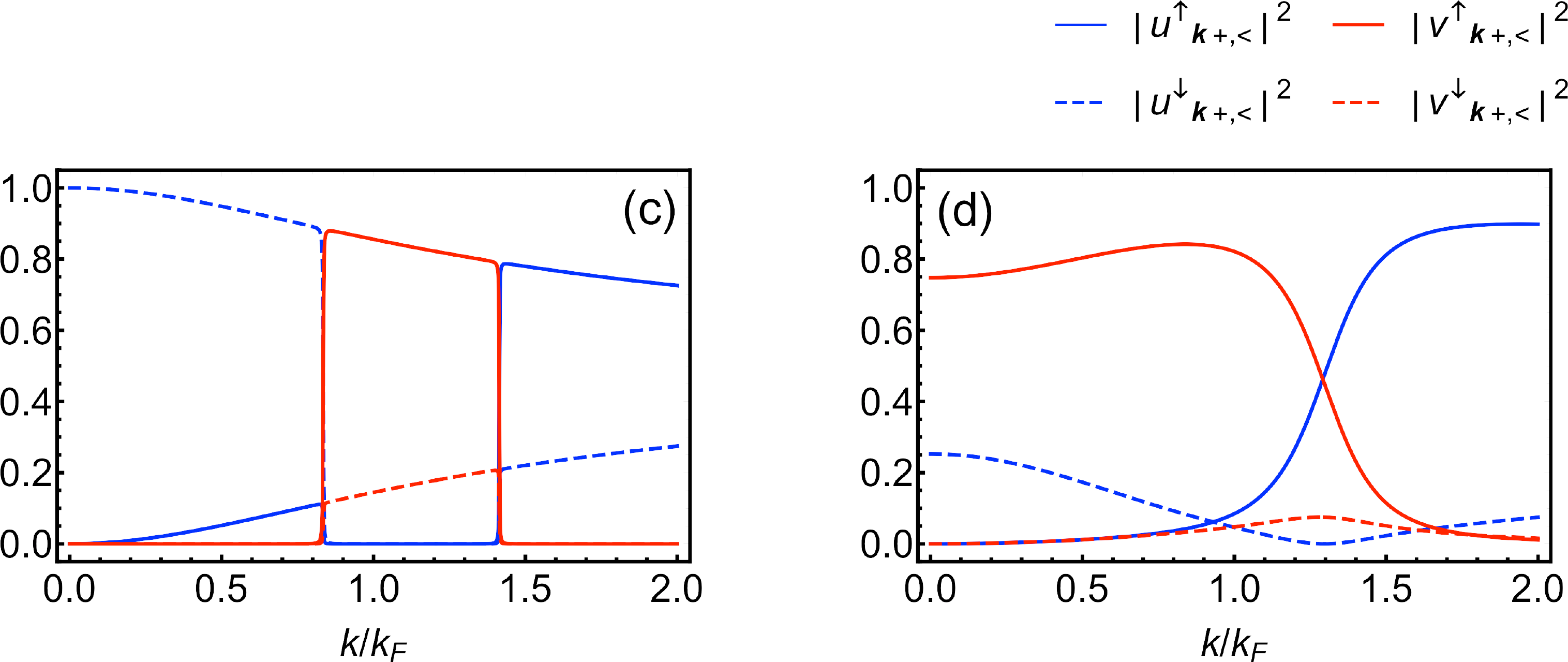}\\[0.2cm]
\hfill\includegraphics[width=0.97\columnwidth]{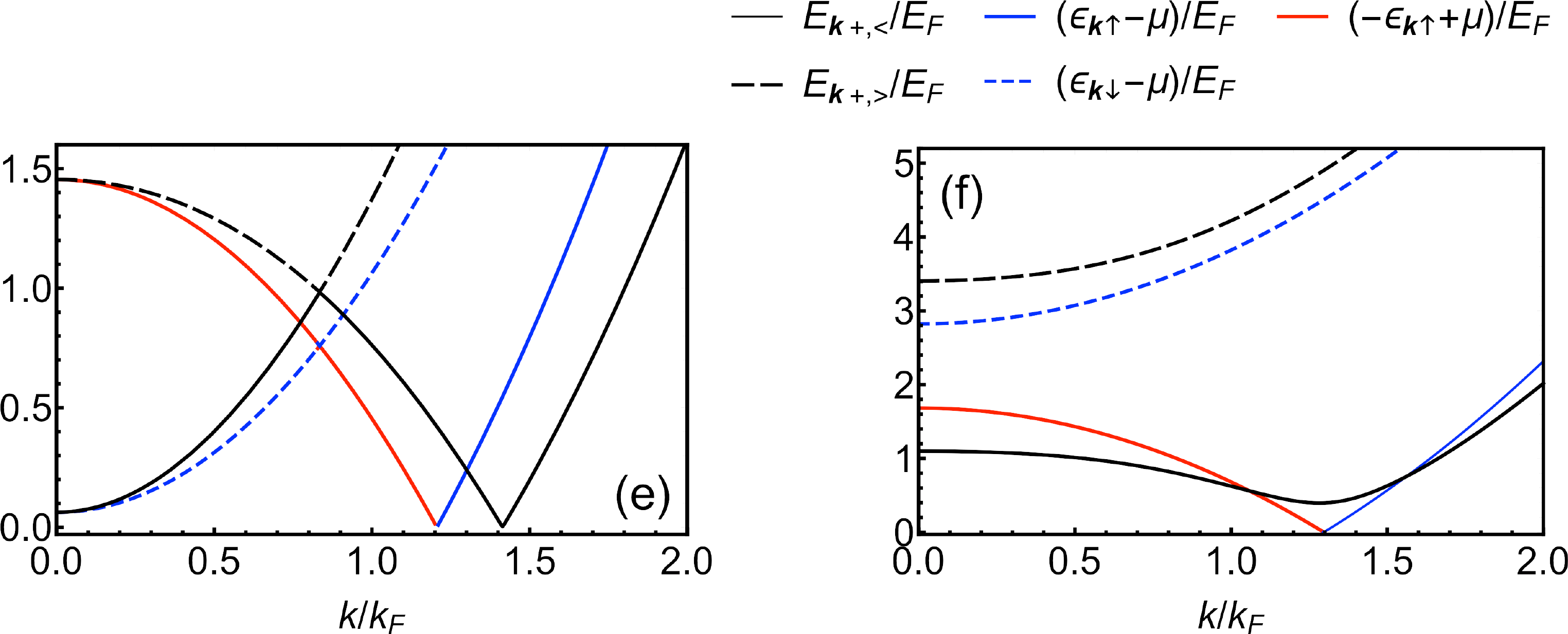}
\caption{\label{fig:topol}%
Canonical features of chiral \textit{p}-wave pairing persist in the
topological-superfluid (TSF) phase of a spin-orbit-coupled 2D Fermi
gas throughout the BCS-to-BEC crossover of the underlying
\textit{s}-wave pairing. Results plotted here have been obtained for
fixed $\lambda k_\mathrm{F}/E_\mathrm{F} = 0.75$, and $k\equiv |\kk|$
is the magnitude of the 2D wave vector of Bogoliubov quasiparticles.
Panels (a), (c) and (e) depict the BCS regime ($E_\mathrm{b}/
E_\mathrm{F}=0.010$ and $h = h_\mathrm{c} + 0.04 \, E_\mathrm{F}$),
whereas panels (b), (d) and (f) are associated with the BEC regime
($E_\mathrm{b}/E_\mathrm{F}=3.0$ and $h = h_>$, corresponding to
the filled square with minimum $h$ in Fig.~\ref{fig:BEClim}). The
momentum-space density distribution $n_k$ has the same distinctive
Fermi-surface feature in both the BCS and BEC regimes of the TSF
[see panels (a) and (b)]. Panel (c) [(d)] shows the spin-resolved
particle and hole probability densities for the lowest positive-energy
branch $E_{\kk +,<}$ of Bogoliubov-quasiparticle excitations whose
energy dispersion is the black solid curve in panel (e) [(f)]. Unlike
in the BEC regime for the NSF [refer to Fig.~\ref{fig:trivial}(d)],
both spin-$\uparrow$-particle and spin-$\uparrow$-hole amplitudes
dominate in the BEC regime of the TSF [depicted here in panel (d)].
The purity of this realization of chiral-\textit{p}-wave pairing
contrasts with the complicated pattern of the spin-resolved Bogoliubov
amplitudes in the BCS regime [panel (c)], which is the result of
small-gap anticrossings between several highly spin-polarized
dispersion branches. For illustration, panels (e) and (f) also show
the purely Zeeman-split dispersions $|\epsilon_{\kk\sigma}-\mu|$.}
\end{figure}

\begin{figure}[t]
\includegraphics[width=0.85\columnwidth]{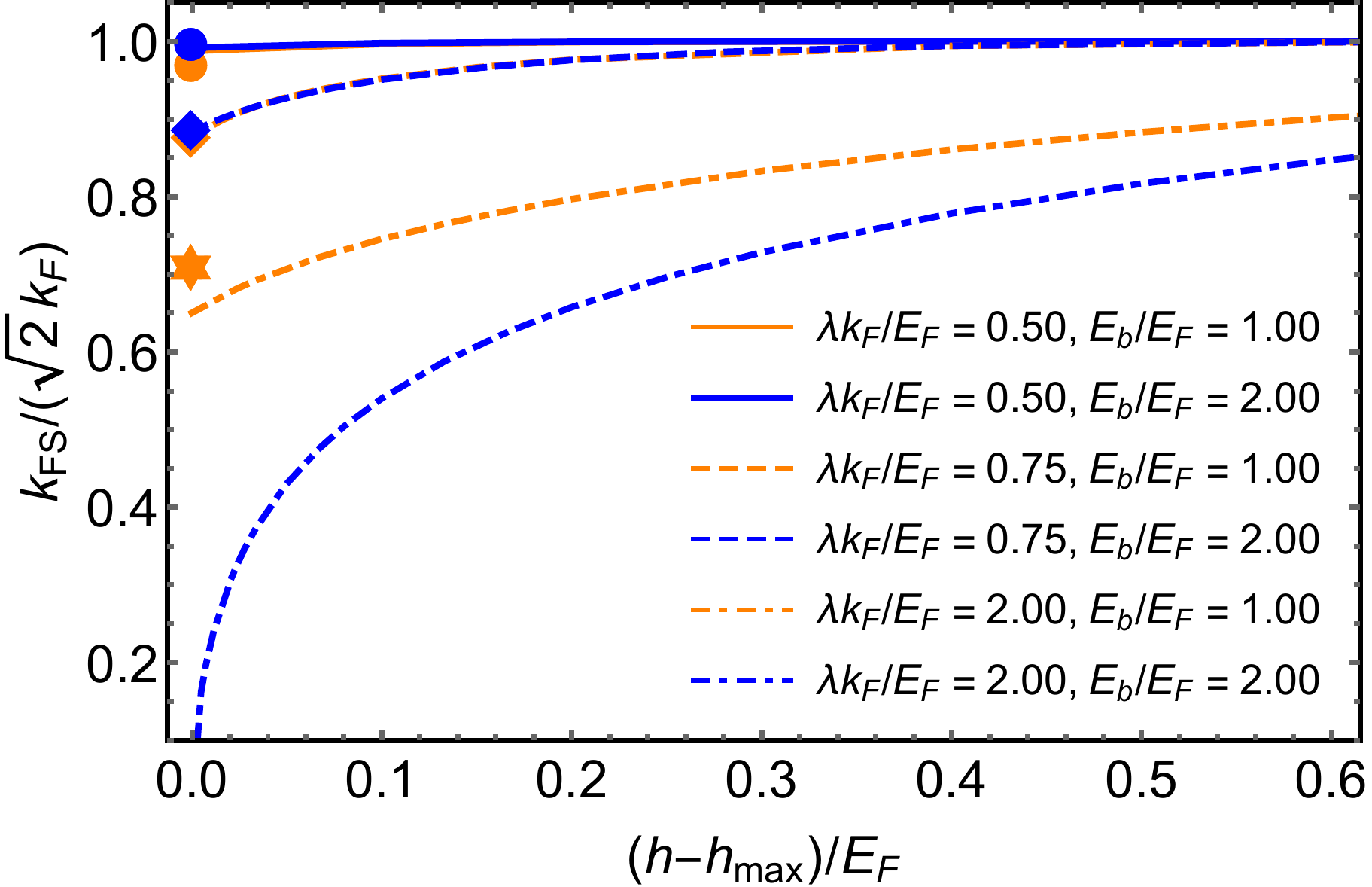}
\caption{\label{fig:kFSvsH}%
Radius $k_\mathrm{FS}$ of the Fermi surface emerging in the
topological-superfluid (TSF) phase of a spin-orbit-coupled 2D Fermi
gas with density $n = m E_\mathrm{F}/(\pi\hbar^2)\equiv k_\mathrm{F}^2
/(2\pi)$ subject to \textit{s}-wave pairing in the BEC regime, defined
formally via the condition (\ref{eq:FSdef}). The parameter $\lambda
k_\mathrm{F}/E_\mathrm{F}$ measures the spin-orbit-coupling strength,
$E_\mathrm{b}$ is the \textit{s}-wave two-particle bound-state energy,
and $h_\mathrm{max}$ corresponds to the critical Zeeman energy above
which the system's ground state is a uniform TSF. The two circles
(diamonds) indicate values predicted for $k_\mathrm{FS}
(h_\mathrm{max})$ by Eq.~(\ref{kFSapprox}), which is expected to be
valid for $\lambda k_\mathrm{F} < h_\mathrm{max}$, using $\lambda
k_\mathrm{F}/E_\mathrm{F} = 0.50$ ($0.75$) and $E_\mathrm{b}/
E_\mathrm{F}$ associated with the correspondingly colored solid
(dashed) line. The orange star indicates the value based on the
estimate $k_\mathrm{FS}(h_\mathrm{max}) \approx \lambda k_\mathrm{F}/
(2 E_\mathrm{F})$ that is expected to apply when $h_\mathrm{max} <
\lambda k_\mathrm{F}$ concomitantly with $\lambda k_\mathrm{F}/
E_\mathrm{F} \gtrsim 1$, which is the case for the parameters of the
orange dot-dashed curve.}
\end{figure}

The typical phenomenology of the BCS-to-BEC crossover for
\textit{s}-wave pairing entails a shift of the dispersion minimum to
$k=0$, Bogoliubov quasi-particles becoming mostly particle-like, and
the momentum-space density distribution loosing its typical
Fermi-surface-like shape~\cite{Nozieres1985,Chen2005,
Astrakharchik2005,Sensarma2007,Shi2015,Shi2016}. This exact scenario
is played out for our more complicated system of interest in the NSF
phase. See Fig.~\ref{fig:trivial} and the extensive discussion in its
caption. In contrast, as illustrated by Fig.~\ref{fig:topol}, all
features associated with effective chiral \textit{p}-wave pairing in
the spin-$\uparrow$ channel remain present throughout the BCS-to-BEC
crossover in the TSF phase. In particular, the momentum-space density
distribution shows a distinctive Fermi-surface edge feature even deep
in the BEC regime for \textit{s}-wave pairing, which is unexpected for
situations where the chemical potential is negative~\footnote{See,
e.g., Ref.~\cite{Chen2005}. Similar behavior to the one found by us
here for the 2D TSF seems to also be implicit in results that were
presented for the 3D spin-orbit-coupled Fermi superfluid  (see, e.g.,
Fig.~6 in Ref.~\cite{Seo2012}) but whose physical significance was
not discussed.}. That the effective \textit{p}-wave pairing retains
BCS-like behavior even as the underlying \textit{s}-wave pairing is
in the BEC regime is illustrated most strikingly by the close
resemblance between the true Bogoliubov-quasiparticle energies and
the dispersions associated with unpaired fermions [see
Fig.~\ref{fig:topol}(f)]. Although much larger generically than in
the BCS regime, $\Delta_\mathrm{pw}$ in the TSF phase for large
$E_\mathrm{b}/E_\mathrm{F}$ is still small enough because of its
dependence on the inverse of the Zeeman energy [see
Eq.~(\ref{eq:pGap})] that the resulting quasiparticle dispersions are
not radically different from those obtained in the absence of pairing.
This contrasts with the NSF phase occurring at lower $h$ for the same
large value of $E_\mathrm{b}/E_\mathrm{F}$ where the unpaired-fermion
dispersions are not at all representative of the lowest-energy branch
of quasiparticle excitations [see Fig.~\ref{fig:trivial}(f)].

The stabilization of the Fermi surface in the TSF phase due to the
larger Zeeman energy is demonstrated in Fig.~\ref{fig:kFSvsH}. Here we
plot the $h$ dependence of the Fermi-surface radius $k_\mathrm{FS}$,
where the latter is defined as the location of the crossing point of
the spin-$\uparrow$-particle and spin-$\uparrow$-hole
Bogoliubov-spinor magnitudes for the lowest-energy quasiparticle
dispersion,
\begin{equation}\label{eq:FSdef}
\big| u^\uparrow_{\kk +, <} \big|^2_{|\kk| = k_\mathrm{FS}} = \big|
v^\uparrow_{\kk +, <} \big|^2_{|\kk| = k_\mathrm{FS}} \quad .
\end{equation}
The condition $|\kk| = k_\mathrm{FS}$ clearly defines a surface in
wave-vector space that separates states having high and low occupation
probabilities, which is the defining property of a Fermi
surface~\cite{Sensarma2007}. We find that a crossing point yielding a
definite value of $k_\mathrm{FS}$ always exists at $h\ge
h_\mathrm{max}$ for any values of $E_\mathrm{b}/E_\mathrm{F}$ and
$\lambda k_\mathrm{F}/E_\mathrm{F}$. For $\lambda k_\mathrm{F}/
E_\mathrm{F} < 1$, the minimum in the dispersion curve $E_{\kk +,<}$
also occurs at $|\kk| = k_\mathrm{FS}$, and the latter's value turns
out to be well-approximated by Eq.~(\ref{kFSapprox}) for $\lambda
k_\mathrm{F} < h_\mathrm{max}$. In situations with very large
spin-orbit coupling $\lambda k_\mathrm{F}/E_\mathrm{F} \gtrsim 1$, the
dispersion minimum is observed to sometimes be absent or appear at
$|\kk| \ne k_\mathrm{FS}$ right after the transition to the TSF phase.
Nevertheless, the coincidence of the quasiparticle-dispersion minimum
and $k_\mathrm{FS}$ is established for $h\gtrsim h_\mathrm{max}$ even
in such cases. Application of the approximate two-band-model results
from Ref.~\onlinecite{Brand2018} to the case $\lambda k_\mathrm{F} >
h_\mathrm{max}$ yields a conservative estimate for the Fermi-surface
radius in this regime, which is given by
\begin{equation}
\frac{k_\mathrm{FS}}{k_\mathrm{F}} = \frac{\lambda k_\mathrm{F}}{2
E_\mathrm{F}} + \sqrt{\frac{\mu}{E_\mathrm{F}} + \frac{\lambda^2
k_\mathrm{F}^2}{4 E_\mathrm{F}^2} + \frac{|\Delta|^2}{E_\mathrm{F}^2}
\left( \frac{E_\mathrm{F}}{\sqrt{2} \lambda k_\mathrm{F}} -
\frac{E_\mathrm{F}}{h} \right)}
\end{equation}
and only holds when the expression under the square-root is positive.
According to results presented in Fig.~\ref{fig:kFSvsH},
$k_\mathrm{FS}$ increases monotonically as a function of $h-
h_\mathrm{max}$ until reaching its asymptotic value $\sqrt{2}\,
k_\mathrm{F}$, which corresponds to the Fermi-surface radius of a
spin-polarized 2D Fermi gas with density $n\equiv k_\mathrm{F}^2/
(2\pi)$.

As can be seen in Fig.~\ref{fig:topol}, the most visible attributes
that distinguish the TSF in the BEC regime from that arising in the
BCS regime are the increased magnitude of the low-energy excitation
gap $\Delta_\mathrm{pw}$ and the strong suppression of the
minority-spin degrees of freedom. The clear dominance of the
spin-$\uparrow$ Bogoliubov amplitudes representing chiral
\textit{p}-wave pairing is one of the favorable qualities exhibited by
the TSF realized in the BEC regime. In addition, a larger magnitude of
$\Delta_\mathrm{pw}$ should help to reduce the influence of many
experimental nonidealities, including thermal fluctuations, as long as
$E_\mathrm{b}/E_\mathrm{F}$ is not too large so that beyond-mean-field
fluctuations have not yet significantly suppressed the value of the
pairing gap. Thus, the TSF realized in the onset of the BEC regime of
the underlying \textit{s}-wave pairing constitutes both a purer and
more-robust version of the highly sought-after chiral \textit{p}-wave
order.

The results obtained and conclusions drawn in our work are based on
the application of mean-field theory. It is well-known that this
method can only provide limited insight into the strongly interacting
(i.e., the BEC) regime of 2D systems~\cite{Kuchiev1996,Bertaina2011,
Salasnich2015,He2015,Shi2015,Shi2016,Turlapov2017}. Here we employed
the mean-field approach to determine (i)~the phase diagram, (ii)~the
magnitude of the pairing gap, and (iii)~momentum-space density
distributions. Before concluding, we discuss the reliability of our
predictions for these three purposes. \textit{(i)~Phase diagrams:\/}
It is generally accepted that zero-temperature phase diagrams obtained
within mean-field theory are qualitatively correct, even in the BEC
regime~\cite{Yi2006,Parish2007,He2008,Fischer2013,Strinati2018}. We
therefore expect the features presented in our work to be similarly
accurate. \textit{(ii)~Pairing-gap magnitude:\/} Suppression of the
pairing gap by beyond-mean-field fluctuations becomes increasingly
important for larger $E_\mathrm{b}$~\cite{He2015,Hu2018}. Therefore,
results for gap magnitudes presented, e.g., in Fig.~\ref{fig:maxDelta}
are only reliable for $E_\mathrm{b}/E_\mathrm{F}\lesssim 1$.
Nevertheless, the conclusion that $E_\mathrm{b}/E_\mathrm{F} \sim 1$
is optimal for realizing a robust TSF continues to hold.
\textit{(iii)~Momentum-space density distributions:\/} Recent
numerical results obtained for our system of interest in the $h=0$
limit (see Supplemental Material for Ref.~\cite{Shi2016}) indicate
that momentum-space density distributions obtained within mean-field
theory are accurate to within $\lesssim 10$\%. Thus our general
conclusions about the re-emergence of a Fermi surface and the
robustness of chiral \textit{p}-wave superfluidity in the BEC regime
of \textit{s}-wave pairing are expected to be valid.

\section{Conclusions and outlook}\label{sec:concl}

We have investigated the strongly interacting regime of the 2D Fermi
gas with \textit{s}-wave pairing, with fixed particle density and
subject to both spin-orbit coupling and Zeeman spin splitting.
Characteristic features of the phase diagram as a function of
two-particle binding energy $E_\mathrm{b}$ and Zeeman energy $h$ are
elucidated and the properties of the homogeneous superfluid phases
studied in greater detail. In particular, we tracked the boundaries of
the homogeneous nontopological and topological superfluids. The
second-order topological-transition line $h_\mathrm{c}\big(
E_\mathrm{b}\big)$, with $h_\mathrm{c}$ defined via
Eq.~(\ref{eq:critZeem}), is truncated by a phase-separation region
that emerges for $E_\mathrm{b}$ larger than a
critical value $E_\mathrm{b}^{(\mathrm{c})}$ that depends on the
spin-orbit-coupling strength (see Figs.~\ref{fig:phaseDia} and
\ref{fig:critEbS}). As a result, the topological transition is of
first order in the limit of large $E_\mathrm{b}/E_\mathrm{F}$.

The homogeneous nontopological phase exhibits all of the expected
features commonly associated with the BCS-to-BEC crossover for
\textit{s}-wave pairing, especially the shrinking, and eventual
disappearance, of an underlying Fermi surface as the Cooper-pair
binding energy is increased. See Fig.~\ref{fig:trivial}(a,b). In
contrast, as illustrated in Fig.~\ref{fig:topol}, the topological
superfluid phase always retains the basic properties of the BCS
regime, including the Fermi-surface characteristics, even for
large $E_\mathrm{b}/E_\mathrm{F}$. This effect demonstrates the
continuity of topological protection through the BCS-to-BEC crossover.
The larger the value of $\lambda k_\mathrm{F}/E_\mathrm{F}$, the
smaller is the Fermi-surface radius $k_\mathrm{FS}$ at the transition
point $h=h_\mathrm{max}$ into the uniform topological phase. With
increasing $h>h_\mathrm{max}$, the Fermi surface is enlarged until its
radius reaches the asymptotic value $\sqrt{2} k_\mathrm{F}$ expected
for a spin-polarized 2D Fermi sea with density $n\equiv k_\mathrm{F}^2
/(2\pi)$. See Fig.~\ref{fig:kFSvsH} for an illustration.

Promising first steps have recently been made towards physical
realization of our system of interest by demonstrating essential
ingredients, e.g., in ultra-cold-atom gases~\cite{Meng2016} and
solid-state heterostructures~\cite{BenShalom2010,Shabani2016}.
State-of-the-art experimental techniques~\cite{Regal2005} could be
utilized, or related theoretical proposals~\cite{Yi2006a} may be
pursued, to confirm the re-appearance of a Fermi surface as the Zeeman
energy is tuned across the topological transition when the system is
in the BEC regime of the underlying \textit{s}-wave pairing. Compared
to the BCS regime, chiral \textit{p}-wave superfluidity realized in
the BEC regime has a larger excitation gap and is less obscured by
minority-spin degrees of freedom, making it the ideal platform for
exploring exotic Majorana excitations in vortices~\cite{Read2000,
Ivanov2001,Gurarie2007} and their potential use for topological
quantum-information-processing paradigms~\cite{DasSarma2015}. Future
work could focus on elucidating also the evolution and properties of
topological superfluids within the phase-separation region.

\begin{acknowledgments}

U.Z.\ thanks W.~Belzig, C.~Bruder, M.~M.~Parish, D.~M.~Stamper-Kurn,
and O.~P.~Sushkov for useful discussions. This work was supported by
the Marsden Fund Council (contract no.\ MAU1604), from NZ government
funding managed by the Royal Society Te Ap\=arangi.

\end{acknowledgments}


\begin{thebibliography}{76}%
\makeatletter
\providecommand \@ifxundefined [1]{%
 \@ifx{#1\undefined}
}%
\providecommand \@ifnum [1]{%
 \ifnum #1\expandafter \@firstoftwo
 \else \expandafter \@secondoftwo
 \fi
}%
\providecommand \@ifx [1]{%
 \ifx #1\expandafter \@firstoftwo
 \else \expandafter \@secondoftwo
 \fi
}%
\providecommand \natexlab [1]{#1}%
\providecommand \enquote  [1]{``#1''}%
\providecommand \bibnamefont  [1]{#1}%
\providecommand \bibfnamefont [1]{#1}%
\providecommand \citenamefont [1]{#1}%
\providecommand \href@noop [0]{\@secondoftwo}%
\providecommand \href [0]{\begingroup \@sanitize@url \@href}%
\providecommand \@href[1]{\@@startlink{#1}\@@href}%
\providecommand \@@href[1]{\endgroup#1\@@endlink}%
\providecommand \@sanitize@url [0]{\catcode `\\12\catcode `\$12\catcode
  `\&12\catcode `\#12\catcode `\^12\catcode `\_12\catcode `\%12\relax}%
\providecommand \@@startlink[1]{}%
\providecommand \@@endlink[0]{}%
\providecommand \url  [0]{\begingroup\@sanitize@url \@url }%
\providecommand \@url [1]{\endgroup\@href {#1}{\urlprefix }}%
\providecommand \urlprefix  [0]{URL }%
\providecommand \Eprint [0]{\href }%
\providecommand \doibase [0]{http://dx.doi.org/}%
\providecommand \selectlanguage [0]{\@gobble}%
\providecommand \bibinfo  [0]{\@secondoftwo}%
\providecommand \bibfield  [0]{\@secondoftwo}%
\providecommand \translation [1]{[#1]}%
\providecommand \BibitemOpen [0]{}%
\providecommand \bibitemStop [0]{}%
\providecommand \bibitemNoStop [0]{.\EOS\space}%
\providecommand \EOS [0]{\spacefactor3000\relax}%
\providecommand \BibitemShut  [1]{\csname bibitem#1\endcsname}%
\let\auto@bib@innerbib\@empty
\bibitem [{\citenamefont {Sato}\ and\ \citenamefont {Ando}(2017)}]{Sato2017}%
  \BibitemOpen
  \bibfield  {author} {\bibinfo {author} {\bibfnamefont {M.}~\bibnamefont
  {Sato}}\ and\ \bibinfo {author} {\bibfnamefont {Y.}~\bibnamefont {Ando}},\
  }\bibfield  {title} {\enquote {\bibinfo {title} {Topological superconductors:
  a review},}\ }\href {\doibase 10.1088/1361-6633/aa6ac7} {\bibfield  {journal}
  {\bibinfo  {journal} {Rep. Prog. Phys.}\ }\textbf {\bibinfo {volume} {80}},\
  \bibinfo {pages} {076501} (\bibinfo {year} {2017})}\BibitemShut {NoStop}%
\bibitem [{\citenamefont {Fu}\ and\ \citenamefont {Kane}(2008)}]{Fu2008}%
  \BibitemOpen
  \bibfield  {author} {\bibinfo {author} {\bibfnamefont {L.}~\bibnamefont
  {Fu}}\ and\ \bibinfo {author} {\bibfnamefont {C.~L.}\ \bibnamefont {Kane}},\
  }\bibfield  {title} {\enquote {\bibinfo {title} {Superconducting proximity
  effect and {Majorana} fermions at the surface of a topological insulator},}\
  }\href {\doibase 10.1103/PhysRevLett.100.096407} {\bibfield  {journal}
  {\bibinfo  {journal} {Phys. Rev. Lett.}\ }\textbf {\bibinfo {volume} {100}},\
  \bibinfo {pages} {096407} (\bibinfo {year} {2008})}\BibitemShut {NoStop}%
\bibitem [{\citenamefont {Zhang}\ \emph {et~al.}(2008)\citenamefont {Zhang},
  \citenamefont {Tewari}, \citenamefont {Lutchyn},\ and\ \citenamefont
  {Das~Sarma}}]{Zhang2008}%
  \BibitemOpen
  \bibfield  {author} {\bibinfo {author} {\bibfnamefont {C.}~\bibnamefont
  {Zhang}}, \bibinfo {author} {\bibfnamefont {S.}~\bibnamefont {Tewari}},
  \bibinfo {author} {\bibfnamefont {R.~M.}\ \bibnamefont {Lutchyn}}, \ and\
  \bibinfo {author} {\bibfnamefont {S.}~\bibnamefont {Das~Sarma}},\ }\bibfield
  {title} {\enquote {\bibinfo {title} {${p}_{x}+i{p}_{y}$ superfluid from
  s-wave interactions of fermionic cold atoms},}\ }\href {\doibase
  10.1103/PhysRevLett.101.160401} {\bibfield  {journal} {\bibinfo  {journal}
  {Phys. Rev. Lett.}\ }\textbf {\bibinfo {volume} {101}},\ \bibinfo {pages}
  {160401} (\bibinfo {year} {2008})}\BibitemShut {NoStop}%
\bibitem [{\citenamefont {Sau}\ \emph {et~al.}(2010)\citenamefont {Sau},
  \citenamefont {Lutchyn}, \citenamefont {Tewari},\ and\ \citenamefont
  {Das~Sarma}}]{Sau2010}%
  \BibitemOpen
  \bibfield  {author} {\bibinfo {author} {\bibfnamefont {J.~D.}\ \bibnamefont
  {Sau}}, \bibinfo {author} {\bibfnamefont {R.~M.}\ \bibnamefont {Lutchyn}},
  \bibinfo {author} {\bibfnamefont {S.}~\bibnamefont {Tewari}}, \ and\ \bibinfo
  {author} {\bibfnamefont {S.}~\bibnamefont {Das~Sarma}},\ }\bibfield  {title}
  {\enquote {\bibinfo {title} {Generic new platform for topological quantum
  computation using semiconductor heterostructures},}\ }\href {\doibase
  10.1103/PhysRevLett.104.040502} {\bibfield  {journal} {\bibinfo  {journal}
  {Phys. Rev. Lett.}\ }\textbf {\bibinfo {volume} {104}},\ \bibinfo {pages}
  {040502} (\bibinfo {year} {2010})}\BibitemShut {NoStop}%
\bibitem [{\citenamefont {Alicea}(2010)}]{Alicea2010}%
  \BibitemOpen
  \bibfield  {author} {\bibinfo {author} {\bibfnamefont {J.}~\bibnamefont
  {Alicea}},\ }\bibfield  {title} {\enquote {\bibinfo {title} {Majorana
  fermions in a tunable semiconductor device},}\ }\href {\doibase
  10.1103/PhysRevB.81.125318} {\bibfield  {journal} {\bibinfo  {journal} {Phys.
  Rev. B}\ }\textbf {\bibinfo {volume} {81}},\ \bibinfo {pages} {125318}
  (\bibinfo {year} {2010})}\BibitemShut {NoStop}%
\bibitem [{\citenamefont {Sato}\ \emph {et~al.}(2010)\citenamefont {Sato},
  \citenamefont {Takahashi},\ and\ \citenamefont {Fujimoto}}]{Sato2010}%
  \BibitemOpen
  \bibfield  {author} {\bibinfo {author} {\bibfnamefont {M.}~\bibnamefont
  {Sato}}, \bibinfo {author} {\bibfnamefont {Y.}~\bibnamefont {Takahashi}}, \
  and\ \bibinfo {author} {\bibfnamefont {S.}~\bibnamefont {Fujimoto}},\
  }\bibfield  {title} {\enquote {\bibinfo {title} {Non-{Abelian} topological
  orders and {Majorana} fermions in spin-singlet superconductors},}\ }\href
  {\doibase 10.1103/PhysRevB.82.134521} {\bibfield  {journal} {\bibinfo
  {journal} {Phys. Rev. B}\ }\textbf {\bibinfo {volume} {82}},\ \bibinfo
  {pages} {134521} (\bibinfo {year} {2010})}\BibitemShut {NoStop}%
\bibitem [{\citenamefont {Chandrasekhar}(1962)}]{Chandrasekhar1962}%
  \BibitemOpen
  \bibfield  {author} {\bibinfo {author} {\bibfnamefont {B.~S.}\ \bibnamefont
  {Chandrasekhar}},\ }\bibfield  {title} {\enquote {\bibinfo {title} {A note on
  the maximum critical field of high-field superconductors},}\ }\href {\doibase
  10.1063/1.1777362} {\bibfield  {journal} {\bibinfo  {journal} {Appl. Phys.
  Lett.}\ }\textbf {\bibinfo {volume} {1}},\ \bibinfo {pages} {7} (\bibinfo
  {year} {1962})}\BibitemShut {NoStop}%
\bibitem [{\citenamefont {Clogston}(1962)}]{Clogston1962}%
  \BibitemOpen
  \bibfield  {author} {\bibinfo {author} {\bibfnamefont {A.~M.}\ \bibnamefont
  {Clogston}},\ }\bibfield  {title} {\enquote {\bibinfo {title} {Upper limit
  for the critical field in hard superconductors},}\ }\href {\doibase
  10.1103/PhysRevLett.9.266} {\bibfield  {journal} {\bibinfo  {journal} {Phys.
  Rev. Lett.}\ }\textbf {\bibinfo {volume} {9}},\ \bibinfo {pages} {266}
  (\bibinfo {year} {1962})}\BibitemShut {NoStop}%
\bibitem [{\citenamefont {He}\ and\ \citenamefont {Zhuang}(2008)}]{He2008}%
  \BibitemOpen
  \bibfield  {author} {\bibinfo {author} {\bibfnamefont {L.}~\bibnamefont
  {He}}\ and\ \bibinfo {author} {\bibfnamefont {P.}~\bibnamefont {Zhuang}},\
  }\bibfield  {title} {\enquote {\bibinfo {title} {Phase diagram of a cold
  polarized {F}ermi gas in two dimensions},}\ }\href {\doibase
  10.1103/PhysRevA.78.033613} {\bibfield  {journal} {\bibinfo  {journal} {Phys.
  Rev. A}\ }\textbf {\bibinfo {volume} {78}},\ \bibinfo {pages} {033613}
  (\bibinfo {year} {2008})}\BibitemShut {NoStop}%
\bibitem [{\citenamefont {Bychkov}\ and\ \citenamefont
  {Rashba}(1984)}]{Bychkov1984}%
  \BibitemOpen
  \bibfield  {author} {\bibinfo {author} {\bibfnamefont {{Yu}.~A.}\
  \bibnamefont {Bychkov}}\ and\ \bibinfo {author} {\bibfnamefont {E.~I.}\
  \bibnamefont {Rashba}},\ }\bibfield  {title} {\enquote {\bibinfo {title}
  {Properties of a {2D} electron gas with lifted spectral degeneracy},}\ }\href
  {http://www.jetpletters.ac.ru/ps/1264/article_19121.shtml} {\bibfield
  {journal} {\bibinfo  {journal} {Pis'ma Zh. Eksp. Teor. Fiz.}\ }\textbf
  {\bibinfo {volume} {39}},\ \bibinfo {pages} {66} (\bibinfo {year} {1984})},\
  \bibinfo {note} {[JETP Lett. \textbf{39}, 78 (1984)]}\BibitemShut {NoStop}%
\bibitem [{\citenamefont {Winkler}(2003)}]{Winkler2003}%
  \BibitemOpen
  \bibfield  {author} {\bibinfo {author} {\bibfnamefont {R.}~\bibnamefont
  {Winkler}},\ }\href {\doibase 10.1007/b13586} {\emph {\bibinfo {title}
  {Spin-Orbit Coupling Effects in Two-Dimensional Electron and Hole Systems}}}\
  (\bibinfo  {publisher} {Springer},\ \bibinfo {address} {Berlin},\ \bibinfo
  {year} {2003})\BibitemShut {NoStop}%
\bibitem [{\citenamefont {Galitski}\ and\ \citenamefont
  {Spielman}(2013)}]{Galitski2013}%
  \BibitemOpen
  \bibfield  {author} {\bibinfo {author} {\bibfnamefont {V.}~\bibnamefont
  {Galitski}}\ and\ \bibinfo {author} {\bibfnamefont {I.~B.}\ \bibnamefont
  {Spielman}},\ }\bibfield  {title} {\enquote {\bibinfo {title} {Spin-orbit
  coupling in quantum gases},}\ }\href {\doibase 10.1038/nature11841}
  {\bibfield  {journal} {\bibinfo  {journal} {Nature (London)}\ }\textbf
  {\bibinfo {volume} {494}},\ \bibinfo {pages} {49} (\bibinfo {year}
  {2013})}\BibitemShut {NoStop}%
\bibitem [{\citenamefont {Brand}\ \emph {et~al.}(2018)\citenamefont {Brand},
  \citenamefont {Toikka},\ and\ \citenamefont {Z{\"u}licke}}]{Brand2018}%
  \BibitemOpen
  \bibfield  {author} {\bibinfo {author} {\bibfnamefont {J.}~\bibnamefont
  {Brand}}, \bibinfo {author} {\bibfnamefont {L.~A.}\ \bibnamefont {Toikka}}, \
  and\ \bibinfo {author} {\bibfnamefont {U.}~\bibnamefont {Z{\"u}licke}},\
  }\bibfield  {title} {\enquote {\bibinfo {title} {Accurate projective two-band
  description of topological superfluidity in spin-orbit-coupled {F}ermi
  gases},}\ }\href {\doibase 10.21468/SciPostPhys.5.2.016} {\bibfield
  {journal} {\bibinfo  {journal} {SciPost Phys.}\ }\textbf {\bibinfo {volume}
  {5}},\ \bibinfo {pages} {16} (\bibinfo {year} {2018})}\BibitemShut {NoStop}%
\bibitem [{\citenamefont {Kallin}\ and\ \citenamefont
  {Berlinsky}(2016)}]{Kallin2016}%
  \BibitemOpen
  \bibfield  {author} {\bibinfo {author} {\bibfnamefont {C.}~\bibnamefont
  {Kallin}}\ and\ \bibinfo {author} {\bibfnamefont {J.}~\bibnamefont
  {Berlinsky}},\ }\bibfield  {title} {\enquote {\bibinfo {title} {Chiral
  superconductors},}\ }\href {\doibase 10.1088/0034-4885/79/5/054502}
  {\bibfield  {journal} {\bibinfo  {journal} {Rep. Prog. Phys.}\ }\textbf
  {\bibinfo {volume} {79}},\ \bibinfo {pages} {054502} (\bibinfo {year}
  {2016})}\BibitemShut {NoStop}%
\bibitem [{\citenamefont {Jackiw}\ and\ \citenamefont
  {Rebbi}(1976)}]{Jackiw1976}%
  \BibitemOpen
  \bibfield  {author} {\bibinfo {author} {\bibfnamefont {R.}~\bibnamefont
  {Jackiw}}\ and\ \bibinfo {author} {\bibfnamefont {C.}~\bibnamefont {Rebbi}},\
  }\bibfield  {title} {\enquote {\bibinfo {title} {Solitons with fermion number
  $\frac{1}{2}$},}\ }\href {\doibase 10.1103/PhysRevD.13.3398} {\bibfield
  {journal} {\bibinfo  {journal} {Phys. Rev. D}\ }\textbf {\bibinfo {volume}
  {13}},\ \bibinfo {pages} {3398--3409} (\bibinfo {year} {1976})}\BibitemShut
  {NoStop}%
\bibitem [{\citenamefont {Kopnin}\ and\ \citenamefont
  {Salomaa}(1991)}]{Kopnin1991}%
  \BibitemOpen
  \bibfield  {author} {\bibinfo {author} {\bibfnamefont {N.~B.}\ \bibnamefont
  {Kopnin}}\ and\ \bibinfo {author} {\bibfnamefont {M.~M.}\ \bibnamefont
  {Salomaa}},\ }\bibfield  {title} {\enquote {\bibinfo {title} {Mutual friction
  in superfluid $^{3}\mathrm{He}$: Effects of bound states in the vortex
  core},}\ }\href {\doibase 10.1103/PhysRevB.44.9667} {\bibfield  {journal}
  {\bibinfo  {journal} {Phys. Rev. B}\ }\textbf {\bibinfo {volume} {44}},\
  \bibinfo {pages} {9667--9677} (\bibinfo {year} {1991})}\BibitemShut {NoStop}%
\bibitem [{\citenamefont {Volovik}(1999)}]{Volovik1999}%
  \BibitemOpen
  \bibfield  {author} {\bibinfo {author} {\bibfnamefont {G.~E.}\ \bibnamefont
  {Volovik}},\ }\bibfield  {title} {\enquote {\bibinfo {title} {Fermion zero
  modes on vortices in chiral superconductors},}\ }\href {\doibase
  10.1134/1.568223} {\bibfield  {journal} {\bibinfo  {journal} {JETP Lett.}\
  }\textbf {\bibinfo {volume} {70}},\ \bibinfo {pages} {609--614} (\bibinfo
  {year} {1999})}\BibitemShut {NoStop}%
\bibitem [{\citenamefont {Read}\ and\ \citenamefont {Green}(2000)}]{Read2000}%
  \BibitemOpen
  \bibfield  {author} {\bibinfo {author} {\bibfnamefont {N.}~\bibnamefont
  {Read}}\ and\ \bibinfo {author} {\bibfnamefont {D.}~\bibnamefont {Green}},\
  }\bibfield  {title} {\enquote {\bibinfo {title} {Paired states of fermions in
  two dimensions with breaking of parity and time-reversal symmetries and the
  fractional quantum {Hall} effect},}\ }\href {\doibase
  10.1103/PhysRevB.61.10267} {\bibfield  {journal} {\bibinfo  {journal} {Phys.
  Rev. B}\ }\textbf {\bibinfo {volume} {61}},\ \bibinfo {pages} {10267--10297}
  (\bibinfo {year} {2000})}\BibitemShut {NoStop}%
\bibitem [{\citenamefont {Ivanov}(2001)}]{Ivanov2001}%
  \BibitemOpen
  \bibfield  {author} {\bibinfo {author} {\bibfnamefont {D.~A.}\ \bibnamefont
  {Ivanov}},\ }\bibfield  {title} {\enquote {\bibinfo {title} {Non-{A}belian
  statistics of half-quantum vortices in $\mathit{p}$-wave superconductors},}\
  }\href {\doibase 10.1103/PhysRevLett.86.268} {\bibfield  {journal} {\bibinfo
  {journal} {Phys. Rev. Lett.}\ }\textbf {\bibinfo {volume} {86}},\ \bibinfo
  {pages} {268--271} (\bibinfo {year} {2001})}\BibitemShut {NoStop}%
\bibitem [{\citenamefont {Gurarie}\ and\ \citenamefont
  {Radzihovsky}(2007)}]{Gurarie2007}%
  \BibitemOpen
  \bibfield  {author} {\bibinfo {author} {\bibfnamefont {V.}~\bibnamefont
  {Gurarie}}\ and\ \bibinfo {author} {\bibfnamefont {L.}~\bibnamefont
  {Radzihovsky}},\ }\bibfield  {title} {\enquote {\bibinfo {title} {Zero modes
  of two-dimensional chiral $p$-wave superconductors},}\ }\href {\doibase
  10.1103/PhysRevB.75.212509} {\bibfield  {journal} {\bibinfo  {journal} {Phys.
  Rev. B}\ }\textbf {\bibinfo {volume} {75}},\ \bibinfo {pages} {212509}
  (\bibinfo {year} {2007})}\BibitemShut {NoStop}%
\bibitem [{\citenamefont {Tewari}\ \emph {et~al.}(2007)\citenamefont {Tewari},
  \citenamefont {{Das Sarma}},\ and\ \citenamefont {Lee}}]{Tewari2007a}%
  \BibitemOpen
  \bibfield  {author} {\bibinfo {author} {\bibfnamefont {S.}~\bibnamefont
  {Tewari}}, \bibinfo {author} {\bibfnamefont {S.}~\bibnamefont {{Das Sarma}}},
  \ and\ \bibinfo {author} {\bibfnamefont {D.-H.}\ \bibnamefont {Lee}},\
  }\bibfield  {title} {\enquote {\bibinfo {title} {Index theorem for the zero
  modes of {M}ajorana fermion vortices in chiral \textit{p}-wave
  superconductors},}\ }\href {\doibase 10.1103/PhysRevLett.99.037001}
  {\bibfield  {journal} {\bibinfo  {journal} {Phys. Rev. Lett.}\ }\textbf
  {\bibinfo {volume} {99}},\ \bibinfo {pages} {037001} (\bibinfo {year}
  {2007})}\BibitemShut {NoStop}%
\bibitem [{\citenamefont {Das~Sarma}\ \emph {et~al.}(2015)\citenamefont
  {Das~Sarma}, \citenamefont {Freedman},\ and\ \citenamefont
  {Nayak}}]{DasSarma2015}%
  \BibitemOpen
  \bibfield  {author} {\bibinfo {author} {\bibfnamefont {S.}~\bibnamefont
  {Das~Sarma}}, \bibinfo {author} {\bibfnamefont {M.}~\bibnamefont {Freedman}},
  \ and\ \bibinfo {author} {\bibfnamefont {C.}~\bibnamefont {Nayak}},\
  }\bibfield  {title} {\enquote {\bibinfo {title} {Majorana zero modes and
  topological quantum computation},}\ }\href {\doibase 10.1038/npjqi.2015.1}
  {\bibfield  {journal} {\bibinfo  {journal} {npj Quant. Inf.}\ }\textbf
  {\bibinfo {volume} {1}},\ \bibinfo {pages} {15001} (\bibinfo {year}
  {2015})}\BibitemShut {NoStop}%
\bibitem [{\citenamefont {Loder}\ \emph {et~al.}(2015)\citenamefont {Loder},
  \citenamefont {Kampf},\ and\ \citenamefont {Kopp}}]{Loder2015}%
  \BibitemOpen
  \bibfield  {author} {\bibinfo {author} {\bibfnamefont {F.}~\bibnamefont
  {Loder}}, \bibinfo {author} {\bibfnamefont {A.~P.}\ \bibnamefont {Kampf}}, \
  and\ \bibinfo {author} {\bibfnamefont {T.}~\bibnamefont {Kopp}},\ }\bibfield
  {title} {\enquote {\bibinfo {title} {Route to topological superconductivity
  via magnetic field rotation},}\ }\href {\doibase 10.1038/srep15302}
  {\bibfield  {journal} {\bibinfo  {journal} {Sci. Rep.}\ }\textbf {\bibinfo
  {volume} {5}},\ \bibinfo {pages} {15302} (\bibinfo {year}
  {2015})}\BibitemShut {NoStop}%
\bibitem [{\citenamefont {Zhai}(2015)}]{Zhai2015}%
  \BibitemOpen
  \bibfield  {author} {\bibinfo {author} {\bibfnamefont {H.}~\bibnamefont
  {Zhai}},\ }\bibfield  {title} {\enquote {\bibinfo {title} {Degenerate quantum
  gases with spin-orbit coupling: a review},}\ }\href {\doibase
  10.1088/0034-4885/78/2/026001} {\bibfield  {journal} {\bibinfo  {journal}
  {Rep. Prog. Phys.}\ }\textbf {\bibinfo {volume} {78}},\ \bibinfo {pages}
  {026001} (\bibinfo {year} {2015})}\BibitemShut {NoStop}%
\bibitem [{\citenamefont {Zhang}\ \emph {et~al.}(2018)\citenamefont {Zhang},
  \citenamefont {Yi},\ and\ \citenamefont {S{\'a}~de Melo}}]{Zhang2018}%
  \BibitemOpen
  \bibinfo {editor} {\bibfnamefont {W.}~\bibnamefont {Zhang}}, \bibinfo
  {editor} {\bibfnamefont {W.}~\bibnamefont {Yi}}, \ and\ \bibinfo {editor}
  {\bibfnamefont {C.~A.~R.}\ \bibnamefont {S{\'a}~de Melo}},\ eds.,\ \href
  {\doibase 10.1142/11050} {\emph {\bibinfo {title} {Synthetic Spin-Orbit
  Coupling in Cold Atoms}}}\ (\bibinfo  {publisher} {World Scientific},\
  \bibinfo {address} {Singapore},\ \bibinfo {year} {2018})\BibitemShut
  {NoStop}%
\bibitem [{\citenamefont {Leggett}(1980{\natexlab{a}})}]{Leggett1980a}%
  \BibitemOpen
  \bibfield  {author} {\bibinfo {author} {\bibfnamefont {A.~J.}\ \bibnamefont
  {Leggett}},\ }\bibfield  {title} {\enquote {\bibinfo {title} {Diatomic
  molecules and {C}ooper pairs},}\ }in\ \href {\doibase 10.1007/BFb0120125}
  {\emph {\bibinfo {booktitle} {Modern Trends in the Theory of Condensed
  Matter}}},\ \bibinfo {editor} {edited by\ \bibinfo {editor} {\bibfnamefont
  {A.}~\bibnamefont {P{\c{e}}kalski}}\ and\ \bibinfo {editor} {\bibfnamefont
  {J.~A.}\ \bibnamefont {Przystawa}}}\ (\bibinfo  {publisher} {Springer},\
  \bibinfo {address} {Berlin},\ \bibinfo {year} {1980})\ pp.\ \bibinfo {pages}
  {13--27}\BibitemShut {NoStop}%
\bibitem [{\citenamefont {Leggett}(1980{\natexlab{b}})}]{Leggett1980b}%
  \BibitemOpen
  \bibfield  {author} {\bibinfo {author} {\bibfnamefont {A.~J.}\ \bibnamefont
  {Leggett}},\ }\bibfield  {title} {\enquote {\bibinfo {title} {Cooper pairing
  in spin-polarized {F}ermi systems},}\ }\href {\doibase
  10.1051/jphyscol:1980704} {\bibfield  {journal} {\bibinfo  {journal} {J.
  Phys. Colloques}\ }\textbf {\bibinfo {volume} {41{\normalfont (C7)}}},\
  \bibinfo {pages} {19--26} (\bibinfo {year} {1980}{\natexlab{b}})}\BibitemShut
  {NoStop}%
\bibitem [{\citenamefont {Randeria}\ \emph {et~al.}(1990)\citenamefont
  {Randeria}, \citenamefont {Duan},\ and\ \citenamefont
  {Shieh}}]{Randeria1990}%
  \BibitemOpen
  \bibfield  {author} {\bibinfo {author} {\bibfnamefont {M.}~\bibnamefont
  {Randeria}}, \bibinfo {author} {\bibfnamefont {J.-M.}\ \bibnamefont {Duan}},
  \ and\ \bibinfo {author} {\bibfnamefont {L.-Y.}\ \bibnamefont {Shieh}},\
  }\bibfield  {title} {\enquote {\bibinfo {title} {Superconductivity in a
  two-dimensional {Fermi} gas: {E}volution from {Cooper} pairing to {Bose}
  condensation},}\ }\href {\doibase 10.1103/PhysRevB.41.327} {\bibfield
  {journal} {\bibinfo  {journal} {Phys. Rev. B}\ }\textbf {\bibinfo {volume}
  {41}},\ \bibinfo {pages} {327--343} (\bibinfo {year} {1990})}\BibitemShut
  {NoStop}%
\bibitem [{\citenamefont {Parish}(2015)}]{Parish2015}%
  \BibitemOpen
  \bibfield  {author} {\bibinfo {author} {\bibfnamefont {M.~M.}\ \bibnamefont
  {Parish}},\ }\bibfield  {title} {\enquote {\bibinfo {title} {The {BCS--BEC}
  crossover},}\ }in\ \href {\doibase 10.1142/9781783264766_0009} {\emph
  {\bibinfo {booktitle} {Quantum Gas Experiments}}},\ \bibinfo {editor} {edited
  by\ \bibinfo {editor} {\bibfnamefont {P.}~\bibnamefont {T\"orm\"a}}\ and\
  \bibinfo {editor} {\bibfnamefont {K.}~\bibnamefont {Sengstock}}}\ (\bibinfo
  {publisher} {Imperial College Press},\ \bibinfo {address} {London},\ \bibinfo
  {year} {2015})\ pp.\ \bibinfo {pages} {179--197}\BibitemShut {NoStop}%
\bibitem [{\citenamefont {Strinati}\ \emph {et~al.}(2018)\citenamefont
  {Strinati}, \citenamefont {Pieri}, \citenamefont {R{\"o}pke}, \citenamefont
  {Schuck},\ and\ \citenamefont {Urban}}]{Strinati2018}%
  \BibitemOpen
  \bibfield  {author} {\bibinfo {author} {\bibfnamefont {G.~C.}\ \bibnamefont
  {Strinati}}, \bibinfo {author} {\bibfnamefont {P.}~\bibnamefont {Pieri}},
  \bibinfo {author} {\bibfnamefont {G.}~\bibnamefont {R{\"o}pke}}, \bibinfo
  {author} {\bibfnamefont {P.}~\bibnamefont {Schuck}}, \ and\ \bibinfo {author}
  {\bibfnamefont {M.}~\bibnamefont {Urban}},\ }\bibfield  {title} {\enquote
  {\bibinfo {title} {The {BCS--BEC} crossover: {F}rom ultra-cold {F}ermi gases
  to nuclear systems},}\ }\href {\doibase 10.1016/j.physrep.2018.02.004}
  {\bibfield  {journal} {\bibinfo  {journal} {Phys. Repts.}\ }\textbf {\bibinfo
  {volume} {738}},\ \bibinfo {pages} {1--76} (\bibinfo {year}
  {2018})}\BibitemShut {NoStop}%
\bibitem [{not()}]{noteEb}%
  \BibitemOpen
  \href@noop {} {}\bibinfo {note} {The energy $E_\mathrm{b}$ measures the
  strength of particle-particle interactions and corresponds to the dimer
  binding energy in the absence of spin-orbit coupling~\cite{Zhai2015}. For
  zero-range interactions, it is related to the 2D \textit{s}-wave scattering
  length $a_\mathrm{2D}$ by $E_\mathrm{b} = \hbar^2/(m\, a_\mathrm{2D})$
  \cite{Bloch2008a,Levinsen2015}, although different conventions for defining
  the scattering length have been adopted in the literature
  \cite{Jeszenszki2018a}.}\BibitemShut {Stop}%
\bibitem [{\citenamefont {Yi}\ and\ \citenamefont {Guo}(2011)}]{Yi2011}%
  \BibitemOpen
  \bibfield  {author} {\bibinfo {author} {\bibfnamefont {W.}~\bibnamefont
  {Yi}}\ and\ \bibinfo {author} {\bibfnamefont {G.-C.}\ \bibnamefont {Guo}},\
  }\bibfield  {title} {\enquote {\bibinfo {title} {Phase separation in a
  polarized {Fermi} gas with spin-orbit coupling},}\ }\href {\doibase
  10.1103/PhysRevA.84.031608} {\bibfield  {journal} {\bibinfo  {journal} {Phys.
  Rev. A}\ }\textbf {\bibinfo {volume} {84}},\ \bibinfo {pages} {031608}
  (\bibinfo {year} {2011})}\BibitemShut {NoStop}%
\bibitem [{\citenamefont {Yang}\ and\ \citenamefont {Wan}(2012)}]{Yang2012}%
  \BibitemOpen
  \bibfield  {author} {\bibinfo {author} {\bibfnamefont {X.}~\bibnamefont
  {Yang}}\ and\ \bibinfo {author} {\bibfnamefont {S.}~\bibnamefont {Wan}},\
  }\bibfield  {title} {\enquote {\bibinfo {title} {Phase diagram of a uniform
  two-dimensional {Fermi} gas with spin-orbit coupling},}\ }\href {\doibase
  10.1103/PhysRevA.85.023633} {\bibfield  {journal} {\bibinfo  {journal} {Phys.
  Rev. A}\ }\textbf {\bibinfo {volume} {85}},\ \bibinfo {pages} {023633}
  (\bibinfo {year} {2012})}\BibitemShut {NoStop}%
\bibitem [{Note1()}]{Note1}%
  \BibitemOpen
  \bibinfo {note} {See, e.g., Fig.~2(a) in Ref.~\cite {He2008}.}\BibitemShut
  {Stop}%
\bibitem [{\citenamefont {Sensarma}\ \emph {et~al.}(2007)\citenamefont
  {Sensarma}, \citenamefont {Randeria},\ and\ \citenamefont
  {Trivedi}}]{Sensarma2007}%
  \BibitemOpen
  \bibfield  {author} {\bibinfo {author} {\bibfnamefont {R.}~\bibnamefont
  {Sensarma}}, \bibinfo {author} {\bibfnamefont {M.}~\bibnamefont {Randeria}},
  \ and\ \bibinfo {author} {\bibfnamefont {N.}~\bibnamefont {Trivedi}},\
  }\bibfield  {title} {\enquote {\bibinfo {title} {Can one determine the
  underlying {F}ermi surface in the superconducting state of strongly
  correlated systems?}}\ }\href {\doibase 10.1103/PhysRevLett.98.027004}
  {\bibfield  {journal} {\bibinfo  {journal} {Phys. Rev. Lett.}\ }\textbf
  {\bibinfo {volume} {98}},\ \bibinfo {pages} {027004} (\bibinfo {year}
  {2007})}\BibitemShut {NoStop}%
\bibitem [{\citenamefont {Nozi{\`e}res}\ and\ \citenamefont
  {Schmitt-Rink}(1985)}]{Nozieres1985}%
  \BibitemOpen
  \bibfield  {author} {\bibinfo {author} {\bibfnamefont {P.}~\bibnamefont
  {Nozi{\`e}res}}\ and\ \bibinfo {author} {\bibfnamefont {S.}~\bibnamefont
  {Schmitt-Rink}},\ }\bibfield  {title} {\enquote {\bibinfo {title} {Bose
  condensation in an attractive fermion gas: {F}rom weak to strong coupling
  superconductivity},}\ }\href {\doibase 10.1007/BF00683774} {\bibfield
  {journal} {\bibinfo  {journal} {J. Low Temp. Phys.}\ }\textbf {\bibinfo
  {volume} {59}},\ \bibinfo {pages} {195--211} (\bibinfo {year}
  {1985})}\BibitemShut {NoStop}%
\bibitem [{\citenamefont {Chen}\ \emph {et~al.}(2005)\citenamefont {Chen},
  \citenamefont {Stajic}, \citenamefont {Tan},\ and\ \citenamefont
  {Levin}}]{Chen2005}%
  \BibitemOpen
  \bibfield  {author} {\bibinfo {author} {\bibfnamefont {Q.}~\bibnamefont
  {Chen}}, \bibinfo {author} {\bibfnamefont {J.}~\bibnamefont {Stajic}},
  \bibinfo {author} {\bibfnamefont {S.}~\bibnamefont {Tan}}, \ and\ \bibinfo
  {author} {\bibfnamefont {K.}~\bibnamefont {Levin}},\ }\bibfield  {title}
  {\enquote {\bibinfo {title} {{BCS--BEC} crossover: {F}rom high temperature
  superconductors to ultracold superfluids},}\ }\href {\doibase
  10.1016/j.physrep.2005.02.005} {\bibfield  {journal} {\bibinfo  {journal}
  {Phys. Rep.}\ }\textbf {\bibinfo {volume} {412}},\ \bibinfo {pages} {1--88}
  (\bibinfo {year} {2005})}\BibitemShut {NoStop}%
\bibitem [{\citenamefont {Astrakharchik}\ \emph {et~al.}(2005)\citenamefont
  {Astrakharchik}, \citenamefont {Boronat}, \citenamefont {Casulleras},\ and\
  \citenamefont {Giorgini}}]{Astrakharchik2005}%
  \BibitemOpen
  \bibfield  {author} {\bibinfo {author} {\bibfnamefont {G.~E.}\ \bibnamefont
  {Astrakharchik}}, \bibinfo {author} {\bibfnamefont {J.}~\bibnamefont
  {Boronat}}, \bibinfo {author} {\bibfnamefont {J.}~\bibnamefont {Casulleras}},
  \ and\ \bibinfo {author} {\bibfnamefont {S.}~\bibnamefont {Giorgini}},\
  }\bibfield  {title} {\enquote {\bibinfo {title} {Momentum distribution and
  condensate fraction of a fermion gas in the {BCS-BEC} crossover},}\ }\href
  {\doibase 10.1103/PhysRevLett.95.230405} {\bibfield  {journal} {\bibinfo
  {journal} {Phys. Rev. Lett.}\ }\textbf {\bibinfo {volume} {95}},\ \bibinfo
  {pages} {230405} (\bibinfo {year} {2005})}\BibitemShut {NoStop}%
\bibitem [{\citenamefont {Shi}\ \emph {et~al.}(2015)\citenamefont {Shi},
  \citenamefont {Chiesa},\ and\ \citenamefont {Zhang}}]{Shi2015}%
  \BibitemOpen
  \bibfield  {author} {\bibinfo {author} {\bibfnamefont {H.}~\bibnamefont
  {Shi}}, \bibinfo {author} {\bibfnamefont {S.}~\bibnamefont {Chiesa}}, \ and\
  \bibinfo {author} {\bibfnamefont {S.}~\bibnamefont {Zhang}},\ }\bibfield
  {title} {\enquote {\bibinfo {title} {Ground-state properties of strongly
  interacting {F}ermi gases in two dimensions},}\ }\href {\doibase
  10.1103/PhysRevA.92.033603} {\bibfield  {journal} {\bibinfo  {journal} {Phys.
  Rev. A}\ }\textbf {\bibinfo {volume} {92}},\ \bibinfo {pages} {033603}
  (\bibinfo {year} {2015})}\BibitemShut {NoStop}%
\bibitem [{\citenamefont {Shi}\ \emph {et~al.}(2016)\citenamefont {Shi},
  \citenamefont {Rosenberg}, \citenamefont {Chiesa},\ and\ \citenamefont
  {Zhang}}]{Shi2016}%
  \BibitemOpen
  \bibfield  {author} {\bibinfo {author} {\bibfnamefont {H.}~\bibnamefont
  {Shi}}, \bibinfo {author} {\bibfnamefont {P.}~\bibnamefont {Rosenberg}},
  \bibinfo {author} {\bibfnamefont {S.}~\bibnamefont {Chiesa}}, \ and\ \bibinfo
  {author} {\bibfnamefont {S.}~\bibnamefont {Zhang}},\ }\bibfield  {title}
  {\enquote {\bibinfo {title} {Rashba spin-orbit coupling, strong interactions,
  and the {BCS-BEC} crossover in the ground state of the two-dimensional
  {F}ermi gas},}\ }\href {\doibase 10.1103/PhysRevLett.117.040401} {\bibfield
  {journal} {\bibinfo  {journal} {Phys. Rev. Lett.}\ }\textbf {\bibinfo
  {volume} {117}},\ \bibinfo {pages} {040401} (\bibinfo {year}
  {2016})}\BibitemShut {NoStop}%
\bibitem [{\citenamefont {Seo}\ \emph {et~al.}(2012)\citenamefont {Seo},
  \citenamefont {Han},\ and\ \citenamefont {S\'a~de Melo}}]{Seo2012}%
  \BibitemOpen
  \bibfield  {author} {\bibinfo {author} {\bibfnamefont {K.}~\bibnamefont
  {Seo}}, \bibinfo {author} {\bibfnamefont {L.}~\bibnamefont {Han}}, \ and\
  \bibinfo {author} {\bibfnamefont {C.~A.~R.}\ \bibnamefont {S\'a~de Melo}},\
  }\bibfield  {title} {\enquote {\bibinfo {title} {Topological phase
  transitions in ultracold {Fermi} superfluids: {The} evolution from
  {Bardeen-Cooper-Schrieffer} to {Bose-Einstein-condensate} superfluids under
  artificial spin-orbit fields},}\ }\href {\doibase 10.1103/PhysRevA.85.033601}
  {\bibfield  {journal} {\bibinfo  {journal} {Phys. Rev. A}\ }\textbf {\bibinfo
  {volume} {85}},\ \bibinfo {pages} {033601} (\bibinfo {year}
  {2012})}\BibitemShut {NoStop}%
\bibitem [{\citenamefont {He}\ and\ \citenamefont {Huang}(2013)}]{He2013}%
  \BibitemOpen
  \bibfield  {author} {\bibinfo {author} {\bibfnamefont {L.}~\bibnamefont
  {He}}\ and\ \bibinfo {author} {\bibfnamefont {X.-G.}\ \bibnamefont {Huang}},\
  }\bibfield  {title} {\enquote {\bibinfo {title} {Superfluidity and collective
  modes in {Rashba} spin-orbit coupled {Fermi} gases},}\ }\href {\doibase
  10.1016/j.aop.2013.06.017} {\bibfield  {journal} {\bibinfo  {journal} {Ann.
  Phys. (Leipzig)}\ }\textbf {\bibinfo {volume} {337}},\ \bibinfo {pages}
  {163--207} (\bibinfo {year} {2013})}\BibitemShut {NoStop}%
\bibitem [{\citenamefont {Yi}\ and\ \citenamefont
  {Duan}(2006{\natexlab{a}})}]{Yi2006}%
  \BibitemOpen
  \bibfield  {author} {\bibinfo {author} {\bibfnamefont {W.}~\bibnamefont
  {Yi}}\ and\ \bibinfo {author} {\bibfnamefont {L.-M.}\ \bibnamefont {Duan}},\
  }\bibfield  {title} {\enquote {\bibinfo {title} {Phase diagram of a polarized
  {F}ermi gas across a {F}eshbach resonance in a potential trap},}\ }\href
  {\doibase 10.1103/PhysRevA.74.013610} {\bibfield  {journal} {\bibinfo
  {journal} {Phys. Rev. A}\ }\textbf {\bibinfo {volume} {74}},\ \bibinfo
  {pages} {013610} (\bibinfo {year} {2006}{\natexlab{a}})}\BibitemShut
  {NoStop}%
\bibitem [{\citenamefont {Parish}\ \emph {et~al.}(2007)\citenamefont {Parish},
  \citenamefont {Marchetti}, \citenamefont {Lamacraft},\ and\ \citenamefont
  {Simons}}]{Parish2007}%
  \BibitemOpen
  \bibfield  {author} {\bibinfo {author} {\bibfnamefont {M.~M.}\ \bibnamefont
  {Parish}}, \bibinfo {author} {\bibfnamefont {F.~M.}\ \bibnamefont
  {Marchetti}}, \bibinfo {author} {\bibfnamefont {A.}~\bibnamefont
  {Lamacraft}}, \ and\ \bibinfo {author} {\bibfnamefont {B.~D.}\ \bibnamefont
  {Simons}},\ }\bibfield  {title} {\enquote {\bibinfo {title}
  {Finite-temperature phase diagram of a polarized {F}ermi condensate},}\
  }\href {\doibase 10.1038/nphys520} {\bibfield  {journal} {\bibinfo  {journal}
  {Nat. Phys.}\ }\textbf {\bibinfo {volume} {3}},\ \bibinfo {pages} {124--128}
  (\bibinfo {year} {2007})}\BibitemShut {NoStop}%
\bibitem [{\citenamefont {Fischer}\ and\ \citenamefont
  {Parish}(2013)}]{Fischer2013}%
  \BibitemOpen
  \bibfield  {author} {\bibinfo {author} {\bibfnamefont {A.~M.}\ \bibnamefont
  {Fischer}}\ and\ \bibinfo {author} {\bibfnamefont {M.~M.}\ \bibnamefont
  {Parish}},\ }\bibfield  {title} {\enquote {\bibinfo {title} {{BCS-BEC}
  crossover in a quasi-two-dimensional {F}ermi gas},}\ }\href {\doibase
  10.1103/PhysRevA.88.023612} {\bibfield  {journal} {\bibinfo  {journal} {Phys.
  Rev. A}\ }\textbf {\bibinfo {volume} {88}},\ \bibinfo {pages} {023612}
  (\bibinfo {year} {2013})}\BibitemShut {NoStop}%
\bibitem [{\citenamefont {Kuchiev}\ and\ \citenamefont
  {Sushkov}(1996)}]{Kuchiev1996}%
  \BibitemOpen
  \bibfield  {author} {\bibinfo {author} {\bibfnamefont {M.~Yu.}\ \bibnamefont
  {Kuchiev}}\ and\ \bibinfo {author} {\bibfnamefont {O.~P.}\ \bibnamefont
  {Sushkov}},\ }\bibfield  {title} {\enquote {\bibinfo {title} {Many-body
  correlation corrections to superconducting pairing in two dimensions},}\
  }\href {\doibase 10.1103/PhysRevB.53.443} {\bibfield  {journal} {\bibinfo
  {journal} {Phys. Rev. B}\ }\textbf {\bibinfo {volume} {53}},\ \bibinfo
  {pages} {443--448} (\bibinfo {year} {1996})}\BibitemShut {NoStop}%
\bibitem [{\citenamefont {Bertaina}\ and\ \citenamefont
  {Giorgini}(2011)}]{Bertaina2011}%
  \BibitemOpen
  \bibfield  {author} {\bibinfo {author} {\bibfnamefont {G.}~\bibnamefont
  {Bertaina}}\ and\ \bibinfo {author} {\bibfnamefont {S.}~\bibnamefont
  {Giorgini}},\ }\bibfield  {title} {\enquote {\bibinfo {title} {{BCS-BEC}
  crossover in a two-dimensional {Fermi} gas},}\ }\href {\doibase
  10.1103/PhysRevLett.106.110403} {\bibfield  {journal} {\bibinfo  {journal}
  {Phys. Rev. Lett.}\ }\textbf {\bibinfo {volume} {106}},\ \bibinfo {pages}
  {110403} (\bibinfo {year} {2011})}\BibitemShut {NoStop}%
\bibitem [{\citenamefont {Salasnich}\ and\ \citenamefont
  {Toigo}(2015)}]{Salasnich2015}%
  \BibitemOpen
  \bibfield  {author} {\bibinfo {author} {\bibfnamefont {L.}~\bibnamefont
  {Salasnich}}\ and\ \bibinfo {author} {\bibfnamefont {F.}~\bibnamefont
  {Toigo}},\ }\bibfield  {title} {\enquote {\bibinfo {title} {Composite bosons
  in the two-dimensional {BCS-BEC} crossover from {Gaussian} fluctuations},}\
  }\href {\doibase 10.1103/PhysRevA.91.011604} {\bibfield  {journal} {\bibinfo
  {journal} {Phys. Rev. A}\ }\textbf {\bibinfo {volume} {91}},\ \bibinfo
  {pages} {011604} (\bibinfo {year} {2015})}\BibitemShut {NoStop}%
\bibitem [{\citenamefont {He}\ \emph {et~al.}(2015)\citenamefont {He},
  \citenamefont {L\"u}, \citenamefont {Cao}, \citenamefont {Hu},\ and\
  \citenamefont {Liu}}]{He2015}%
  \BibitemOpen
  \bibfield  {author} {\bibinfo {author} {\bibfnamefont {L.}~\bibnamefont
  {He}}, \bibinfo {author} {\bibfnamefont {H.}~\bibnamefont {L\"u}}, \bibinfo
  {author} {\bibfnamefont {G.}~\bibnamefont {Cao}}, \bibinfo {author}
  {\bibfnamefont {H.}~\bibnamefont {Hu}}, \ and\ \bibinfo {author}
  {\bibfnamefont {X.-J.}\ \bibnamefont {Liu}},\ }\bibfield  {title} {\enquote
  {\bibinfo {title} {Quantum fluctuations in the {BCS-BEC} crossover of
  two-dimensional {F}ermi gases},}\ }\href {\doibase
  10.1103/PhysRevA.92.023620} {\bibfield  {journal} {\bibinfo  {journal} {Phys.
  Rev. A}\ }\textbf {\bibinfo {volume} {92}},\ \bibinfo {pages} {023620}
  (\bibinfo {year} {2015})}\BibitemShut {NoStop}%
\bibitem [{\citenamefont {Turlapov}\ and\ \citenamefont
  {Kagan}(2017)}]{Turlapov2017}%
  \BibitemOpen
  \bibfield  {author} {\bibinfo {author} {\bibfnamefont {A.~V.}\ \bibnamefont
  {Turlapov}}\ and\ \bibinfo {author} {\bibfnamefont {M.~{Yu}.}\ \bibnamefont
  {Kagan}},\ }\bibfield  {title} {\enquote {\bibinfo {title} {{Fermi-to-Bose}
  crossover in a trapped quasi-{2D} gas of fermionic atoms},}\ }\href {\doibase
  10.1088/1361-648x/aa7ad9} {\bibfield  {journal} {\bibinfo  {journal} {J.
  Phys.: Condens. Matter}\ }\textbf {\bibinfo {volume} {29}},\ \bibinfo {pages}
  {383004} (\bibinfo {year} {2017})}\BibitemShut {NoStop}%
\bibitem [{\citenamefont {Hu}\ \emph {et~al.}(2018)\citenamefont {Hu},
  \citenamefont {Mulkerin}, \citenamefont {He}, \citenamefont {Wang},\ and\
  \citenamefont {Liu}}]{Hu2018}%
  \BibitemOpen
  \bibfield  {author} {\bibinfo {author} {\bibfnamefont {H.}~\bibnamefont
  {Hu}}, \bibinfo {author} {\bibfnamefont {B.~C.}\ \bibnamefont {Mulkerin}},
  \bibinfo {author} {\bibfnamefont {L.}~\bibnamefont {He}}, \bibinfo {author}
  {\bibfnamefont {J.}~\bibnamefont {Wang}}, \ and\ \bibinfo {author}
  {\bibfnamefont {X.-J.}\ \bibnamefont {Liu}},\ }\bibfield  {title} {\enquote
  {\bibinfo {title} {Quantum fluctuations of a resonantly interacting $p$-wave
  {Fermi} superfluid in two dimensions},}\ }\href {\doibase
  10.1103/PhysRevA.98.063605} {\bibfield  {journal} {\bibinfo  {journal} {Phys.
  Rev. A}\ }\textbf {\bibinfo {volume} {98}},\ \bibinfo {pages} {063605}
  (\bibinfo {year} {2018})}\BibitemShut {NoStop}%
\bibitem [{\citenamefont {de~Gennes}(1989)}]{deGennes1989}%
  \BibitemOpen
  \bibfield  {author} {\bibinfo {author} {\bibfnamefont {P.~G.}\ \bibnamefont
  {de~Gennes}},\ }\href@noop {} {\emph {\bibinfo {title} {Superconductivity of
  Metals and Alloys}}}\ (\bibinfo  {publisher} {Addison-Wesley},\ \bibinfo
  {address} {Reading, MA},\ \bibinfo {year} {1989})\BibitemShut {NoStop}%
\bibitem [{Note2()}]{Note2}%
  \BibitemOpen
  \bibinfo {note} {Our notation adheres to that used in Ref.~\cite
  {Brand2018}.}\BibitemShut {Stop}%
\bibitem [{Note3()}]{Note3}%
  \BibitemOpen
  \bibinfo {note} {While we adopt the 2D-Rashba form~\cite {Bychkov1984} for
  $\lambda _{\protect \ensuremath {\protect \bm {\protect \mathrm {k}}}}$, our
  results apply also to other types of spin-orbit coupling that depend linearly
  on the components of ${\protect \ensuremath {\protect \bm {\protect \mathrm
  {k}}}}$, such as the 2D-Dirac and 2D-Dresselhaus functional forms~\cite
  {Winkler2003,Galitski2013} corresponding to $\lambda _{\protect \ensuremath
  {\protect \bm {\protect \mathrm {k}}}}=\lambda (k_x-i k_y)$ and $\lambda (k_x
  + i k_y)$, respectively.}\BibitemShut {Stop}%
\bibitem [{\citenamefont {Zhou}\ \emph {et~al.}(2011)\citenamefont {Zhou},
  \citenamefont {Zhang},\ and\ \citenamefont {Yi}}]{Zhou2011}%
  \BibitemOpen
  \bibfield  {author} {\bibinfo {author} {\bibfnamefont {J.}~\bibnamefont
  {Zhou}}, \bibinfo {author} {\bibfnamefont {W.}~\bibnamefont {Zhang}}, \ and\
  \bibinfo {author} {\bibfnamefont {W.}~\bibnamefont {Yi}},\ }\bibfield
  {title} {\enquote {\bibinfo {title} {Topological superfluid in a trapped
  two-dimensional polarized {Fermi} gas with spin-orbit coupling},}\ }\href
  {\doibase 10.1103/PhysRevA.84.063603} {\bibfield  {journal} {\bibinfo
  {journal} {Phys. Rev. A}\ }\textbf {\bibinfo {volume} {84}},\ \bibinfo
  {pages} {063603} (\bibinfo {year} {2011})}\BibitemShut {NoStop}%
\bibitem [{\citenamefont {Radzihovsky}\ and\ \citenamefont
  {Sheehy}(2010)}]{Radzihovsky2010}%
  \BibitemOpen
  \bibfield  {author} {\bibinfo {author} {\bibfnamefont {L.}~\bibnamefont
  {Radzihovsky}}\ and\ \bibinfo {author} {\bibfnamefont {D.~E.}\ \bibnamefont
  {Sheehy}},\ }\bibfield  {title} {\enquote {\bibinfo {title} {Imbalanced
  {F}eshbach-resonant {F}ermi gases},}\ }\href {\doibase
  10.1088/0034-4885/73/7/076501} {\bibfield  {journal} {\bibinfo  {journal}
  {Rep. Prog. Phys.}\ }\textbf {\bibinfo {volume} {73}},\ \bibinfo {pages}
  {076501} (\bibinfo {year} {2010})}\BibitemShut {NoStop}%
\bibitem [{\citenamefont {Sheehy}\ and\ \citenamefont
  {Radzihovsky}(2007{\natexlab{a}})}]{Sheehy2007}%
  \BibitemOpen
  \bibfield  {author} {\bibinfo {author} {\bibfnamefont {D.~E.}\ \bibnamefont
  {Sheehy}}\ and\ \bibinfo {author} {\bibfnamefont {L.}~\bibnamefont
  {Radzihovsky}},\ }\bibfield  {title} {\enquote {\bibinfo {title} {{BEC--BCS}
  crossover, phase transitions and phase separation in polarized
  resonantly-paired superfluids},}\ }\href {\doibase 10.1016/j.aop.2006.09.009}
  {\bibfield  {journal} {\bibinfo  {journal} {Ann. Phys. (NY)}\ }\textbf
  {\bibinfo {volume} {322}},\ \bibinfo {pages} {1790--1924} (\bibinfo {year}
  {2007}{\natexlab{a}})}\BibitemShut {NoStop}%
\bibitem [{\citenamefont {Sheehy}\ and\ \citenamefont
  {Radzihovsky}(2007{\natexlab{b}})}]{Sheehy2007a}%
  \BibitemOpen
  \bibfield  {author} {\bibinfo {author} {\bibfnamefont {D.~E.}\ \bibnamefont
  {Sheehy}}\ and\ \bibinfo {author} {\bibfnamefont {L.}~\bibnamefont
  {Radzihovsky}},\ }\bibfield  {title} {\enquote {\bibinfo {title} {Comment on
  ``{S}uperfluid stability in the {BEC-BCS} crossover''},}\ }\href {\doibase
  10.1103/PhysRevB.75.136501} {\bibfield  {journal} {\bibinfo  {journal} {Phys.
  Rev. B}\ }\textbf {\bibinfo {volume} {75}},\ \bibinfo {pages} {136501}
  (\bibinfo {year} {2007}{\natexlab{b}})}\BibitemShut {NoStop}%
\bibitem [{\citenamefont {Sarma}(1963)}]{Sarma1963}%
  \BibitemOpen
  \bibfield  {author} {\bibinfo {author} {\bibfnamefont {G.}~\bibnamefont
  {Sarma}},\ }\bibfield  {title} {\enquote {\bibinfo {title} {On the influence
  of a uniform exchange field acting on the spins of the conduction electrons
  in a superconductor},}\ }\href {\doibase 10.1016/0022-3697(63)90007-6}
  {\bibfield  {journal} {\bibinfo  {journal} {J. Phys. Chem. Solids}\ }\textbf
  {\bibinfo {volume} {24}},\ \bibinfo {pages} {1029--1032} (\bibinfo {year}
  {1963})}\BibitemShut {NoStop}%
\bibitem [{\citenamefont {Lamacraft}\ and\ \citenamefont
  {Marchetti}(2008)}]{Lamacraft2008}%
  \BibitemOpen
  \bibfield  {author} {\bibinfo {author} {\bibfnamefont {A.}~\bibnamefont
  {Lamacraft}}\ and\ \bibinfo {author} {\bibfnamefont {F.~M.}\ \bibnamefont
  {Marchetti}},\ }\bibfield  {title} {\enquote {\bibinfo {title} {Spinodal
  decomposition in polarized {F}ermi superfluids},}\ }\href {\doibase
  10.1103/PhysRevB.77.014511} {\bibfield  {journal} {\bibinfo  {journal} {Phys.
  Rev. B}\ }\textbf {\bibinfo {volume} {77}},\ \bibinfo {pages} {014511}
  (\bibinfo {year} {2008})}\BibitemShut {NoStop}%
\bibitem [{\citenamefont {He}\ and\ \citenamefont {Zhuang}(2009)}]{He2009}%
  \BibitemOpen
  \bibfield  {author} {\bibinfo {author} {\bibfnamefont {L.}~\bibnamefont
  {He}}\ and\ \bibinfo {author} {\bibfnamefont {P.}~\bibnamefont {Zhuang}},\
  }\bibfield  {title} {\enquote {\bibinfo {title} {Stable {S}arma state in
  two-band {F}ermi systems},}\ }\href {\doibase 10.1103/PhysRevB.79.024511}
  {\bibfield  {journal} {\bibinfo  {journal} {Phys. Rev. B}\ }\textbf {\bibinfo
  {volume} {79}},\ \bibinfo {pages} {024511} (\bibinfo {year}
  {2009})}\BibitemShut {NoStop}%
\bibitem [{\citenamefont {Bedaque}\ \emph {et~al.}(2003)\citenamefont
  {Bedaque}, \citenamefont {Caldas},\ and\ \citenamefont
  {Rupak}}]{Bedaque2003}%
  \BibitemOpen
  \bibfield  {author} {\bibinfo {author} {\bibfnamefont {P.~F.}\ \bibnamefont
  {Bedaque}}, \bibinfo {author} {\bibfnamefont {H.}~\bibnamefont {Caldas}}, \
  and\ \bibinfo {author} {\bibfnamefont {G.}~\bibnamefont {Rupak}},\ }\bibfield
   {title} {\enquote {\bibinfo {title} {Phase separation in asymmetrical
  fermion superfluids},}\ }\href {\doibase 10.1103/PhysRevLett.91.247002}
  {\bibfield  {journal} {\bibinfo  {journal} {Phys. Rev. Lett.}\ }\textbf
  {\bibinfo {volume} {91}},\ \bibinfo {pages} {247002} (\bibinfo {year}
  {2003})}\BibitemShut {NoStop}%
\bibitem [{\citenamefont {Carlson}\ and\ \citenamefont
  {Reddy}(2005)}]{Carlson2005}%
  \BibitemOpen
  \bibfield  {author} {\bibinfo {author} {\bibfnamefont {J.}~\bibnamefont
  {Carlson}}\ and\ \bibinfo {author} {\bibfnamefont {S.}~\bibnamefont
  {Reddy}},\ }\bibfield  {title} {\enquote {\bibinfo {title} {Asymmetric
  two-component fermion systems in strong coupling},}\ }\href {\doibase
  10.1103/PhysRevLett.95.060401} {\bibfield  {journal} {\bibinfo  {journal}
  {Phys. Rev. Lett.}\ }\textbf {\bibinfo {volume} {95}},\ \bibinfo {pages}
  {060401} (\bibinfo {year} {2005})}\BibitemShut {NoStop}%
\bibitem [{\citenamefont {Sheehy}\ and\ \citenamefont
  {Radzihovsky}(2006)}]{Sheehy2006}%
  \BibitemOpen
  \bibfield  {author} {\bibinfo {author} {\bibfnamefont {D.~E.}\ \bibnamefont
  {Sheehy}}\ and\ \bibinfo {author} {\bibfnamefont {L.}~\bibnamefont
  {Radzihovsky}},\ }\bibfield  {title} {\enquote {\bibinfo {title} {{BEC-BCS}
  crossover in ``magnetized'' {F}eshbach-resonantly paired superfluids},}\
  }\href {\doibase 10.1103/PhysRevLett.96.060401} {\bibfield  {journal}
  {\bibinfo  {journal} {Phys. Rev. Lett.}\ }\textbf {\bibinfo {volume} {96}},\
  \bibinfo {pages} {060401} (\bibinfo {year} {2006})}\BibitemShut {NoStop}%
\bibitem [{\citenamefont {Son}\ and\ \citenamefont
  {Stephanov}(2006)}]{Son2006}%
  \BibitemOpen
  \bibfield  {author} {\bibinfo {author} {\bibfnamefont {D.~T.}\ \bibnamefont
  {Son}}\ and\ \bibinfo {author} {\bibfnamefont {M.~A.}\ \bibnamefont
  {Stephanov}},\ }\bibfield  {title} {\enquote {\bibinfo {title} {Phase diagram
  of a cold polarized {F}ermi gas},}\ }\href {\doibase
  10.1103/PhysRevA.74.013614} {\bibfield  {journal} {\bibinfo  {journal} {Phys.
  Rev. A}\ }\textbf {\bibinfo {volume} {74}},\ \bibinfo {pages} {013614}
  (\bibinfo {year} {2006})}\BibitemShut {NoStop}%
\bibitem [{\citenamefont {Du}\ \emph {et~al.}(2009)\citenamefont {Du},
  \citenamefont {Chen},\ and\ \citenamefont {Liang}}]{Du2009}%
  \BibitemOpen
  \bibfield  {author} {\bibinfo {author} {\bibfnamefont {J.-J.}\ \bibnamefont
  {Du}}, \bibinfo {author} {\bibfnamefont {C.}~\bibnamefont {Chen}}, \ and\
  \bibinfo {author} {\bibfnamefont {J.-J.}\ \bibnamefont {Liang}},\ }\bibfield
  {title} {\enquote {\bibinfo {title} {Asymmetric two-component {F}ermi gas in
  two dimensions},}\ }\href {\doibase 10.1103/PhysRevA.80.023601} {\bibfield
  {journal} {\bibinfo  {journal} {Phys. Rev. A}\ }\textbf {\bibinfo {volume}
  {80}},\ \bibinfo {pages} {023601} (\bibinfo {year} {2009})}\BibitemShut
  {NoStop}%
\bibitem [{Note4()}]{Note4}%
  \BibitemOpen
  \bibinfo {note} {This task becomes particularly challenging for the part of
  the phase diagram where multiple selfconsistent solutions of the gap equation
  exist at fixed $h$. Generally, two of these correspond to minima of the
  ground-state energy taken at fixed $\mu $, and their combined evolution
  between global- or local-minimum status needs to be tracked.}\BibitemShut
  {Stop}%
\bibitem [{Note5()}]{Note5}%
  \BibitemOpen
  \bibinfo {note} {See, e.g., Ref.~\cite {Chen2005}. Similar behavior to the
  one found by us here for the 2D TSF seems to also be implicit in results that
  were presented for the 3D spin-orbit-coupled Fermi superfluid (see, e.g.,
  Fig.~6 in Ref.~\cite {Seo2012}) but whose physical significance was not
  discussed.}\BibitemShut {Stop}%
\bibitem [{\citenamefont {Meng}\ \emph {et~al.}(2016)\citenamefont {Meng},
  \citenamefont {Huang}, \citenamefont {Peng}, \citenamefont {Li},
  \citenamefont {Chen}, \citenamefont {Xu}, \citenamefont {Zhang},
  \citenamefont {Wang},\ and\ \citenamefont {Zhang}}]{Meng2016}%
  \BibitemOpen
  \bibfield  {author} {\bibinfo {author} {\bibfnamefont {Z.}~\bibnamefont
  {Meng}}, \bibinfo {author} {\bibfnamefont {L.}~\bibnamefont {Huang}},
  \bibinfo {author} {\bibfnamefont {P.}~\bibnamefont {Peng}}, \bibinfo {author}
  {\bibfnamefont {D.}~\bibnamefont {Li}}, \bibinfo {author} {\bibfnamefont
  {L.}~\bibnamefont {Chen}}, \bibinfo {author} {\bibfnamefont {Y.}~\bibnamefont
  {Xu}}, \bibinfo {author} {\bibfnamefont {C.}~\bibnamefont {Zhang}}, \bibinfo
  {author} {\bibfnamefont {P.}~\bibnamefont {Wang}}, \ and\ \bibinfo {author}
  {\bibfnamefont {J.}~\bibnamefont {Zhang}},\ }\bibfield  {title} {\enquote
  {\bibinfo {title} {Experimental observation of a topological band gap opening
  in ultracold {Fermi} gases with two-dimensional spin-orbit coupling},}\
  }\href {\doibase 10.1103/PhysRevLett.117.235304} {\bibfield  {journal}
  {\bibinfo  {journal} {Phys. Rev. Lett.}\ }\textbf {\bibinfo {volume} {117}},\
  \bibinfo {pages} {235304} (\bibinfo {year} {2016})}\BibitemShut {NoStop}%
\bibitem [{\citenamefont {Ben~Shalom}\ \emph {et~al.}(2010)\citenamefont
  {Ben~Shalom}, \citenamefont {Sachs}, \citenamefont {Rakhmilevitch},
  \citenamefont {Palevski},\ and\ \citenamefont {Dagan}}]{BenShalom2010}%
  \BibitemOpen
  \bibfield  {author} {\bibinfo {author} {\bibfnamefont {M.}~\bibnamefont
  {Ben~Shalom}}, \bibinfo {author} {\bibfnamefont {M.}~\bibnamefont {Sachs}},
  \bibinfo {author} {\bibfnamefont {D.}~\bibnamefont {Rakhmilevitch}}, \bibinfo
  {author} {\bibfnamefont {A.}~\bibnamefont {Palevski}}, \ and\ \bibinfo
  {author} {\bibfnamefont {Y.}~\bibnamefont {Dagan}},\ }\bibfield  {title}
  {\enquote {\bibinfo {title} {Tuning spin-orbit coupling and superconductivity
  at the {SrTiO}$_{3}$/{LaAlO}$_{3}$ interface: {A} magnetotransport study},}\
  }\href {\doibase 10.1103/PhysRevLett.104.126802} {\bibfield  {journal}
  {\bibinfo  {journal} {Phys. Rev. Lett.}\ }\textbf {\bibinfo {volume} {104}},\
  \bibinfo {pages} {126802} (\bibinfo {year} {2010})}\BibitemShut {NoStop}%
\bibitem [{\citenamefont {Shabani}\ \emph {et~al.}(2016)\citenamefont
  {Shabani}, \citenamefont {Kjaergaard}, \citenamefont {Suominen},
  \citenamefont {Kim}, \citenamefont {Nichele}, \citenamefont {Pakrouski},
  \citenamefont {Stankevic}, \citenamefont {Lutchyn}, \citenamefont
  {Krogstrup}, \citenamefont {Feidenhans'l}, \citenamefont {Kraemer},
  \citenamefont {Nayak}, \citenamefont {Troyer}, \citenamefont {Marcus},\ and\
  \citenamefont {Palmstr\o{}m}}]{Shabani2016}%
  \BibitemOpen
  \bibfield  {author} {\bibinfo {author} {\bibfnamefont {J.}~\bibnamefont
  {Shabani}}, \bibinfo {author} {\bibfnamefont {M.}~\bibnamefont {Kjaergaard}},
  \bibinfo {author} {\bibfnamefont {H.~J.}\ \bibnamefont {Suominen}}, \bibinfo
  {author} {\bibfnamefont {Y.}~\bibnamefont {Kim}}, \bibinfo {author}
  {\bibfnamefont {F.}~\bibnamefont {Nichele}}, \bibinfo {author} {\bibfnamefont
  {K.}~\bibnamefont {Pakrouski}}, \bibinfo {author} {\bibfnamefont
  {T.}~\bibnamefont {Stankevic}}, \bibinfo {author} {\bibfnamefont {R.~M.}\
  \bibnamefont {Lutchyn}}, \bibinfo {author} {\bibfnamefont {P.}~\bibnamefont
  {Krogstrup}}, \bibinfo {author} {\bibfnamefont {R.}~\bibnamefont
  {Feidenhans'l}}, \bibinfo {author} {\bibfnamefont {S.}~\bibnamefont
  {Kraemer}}, \bibinfo {author} {\bibfnamefont {C.}~\bibnamefont {Nayak}},
  \bibinfo {author} {\bibfnamefont {M.}~\bibnamefont {Troyer}}, \bibinfo
  {author} {\bibfnamefont {C.~M.}\ \bibnamefont {Marcus}}, \ and\ \bibinfo
  {author} {\bibfnamefont {C.~J.}\ \bibnamefont {Palmstr\o{}m}},\ }\bibfield
  {title} {\enquote {\bibinfo {title} {Two-dimensional epitaxial
  superconductor-semiconductor heterostructures: {A} platform for topological
  superconducting networks},}\ }\href {\doibase 10.1103/PhysRevB.93.155402}
  {\bibfield  {journal} {\bibinfo  {journal} {Phys. Rev. B}\ }\textbf {\bibinfo
  {volume} {93}},\ \bibinfo {pages} {155402} (\bibinfo {year}
  {2016})}\BibitemShut {NoStop}%
\bibitem [{\citenamefont {Regal}\ \emph {et~al.}(2005)\citenamefont {Regal},
  \citenamefont {Greiner}, \citenamefont {Giorgini}, \citenamefont {Holland},\
  and\ \citenamefont {Jin}}]{Regal2005}%
  \BibitemOpen
  \bibfield  {author} {\bibinfo {author} {\bibfnamefont {C.~A.}\ \bibnamefont
  {Regal}}, \bibinfo {author} {\bibfnamefont {M.}~\bibnamefont {Greiner}},
  \bibinfo {author} {\bibfnamefont {S.}~\bibnamefont {Giorgini}}, \bibinfo
  {author} {\bibfnamefont {M.}~\bibnamefont {Holland}}, \ and\ \bibinfo
  {author} {\bibfnamefont {D.~S.}\ \bibnamefont {Jin}},\ }\bibfield  {title}
  {\enquote {\bibinfo {title} {Momentum distribution of a {F}ermi gas of atoms
  in the {BCS-BEC} crossover},}\ }\href {\doibase
  10.1103/PhysRevLett.95.250404} {\bibfield  {journal} {\bibinfo  {journal}
  {Phys. Rev. Lett.}\ }\textbf {\bibinfo {volume} {95}},\ \bibinfo {pages}
  {250404} (\bibinfo {year} {2005})}\BibitemShut {NoStop}%
\bibitem [{\citenamefont {Yi}\ and\ \citenamefont
  {Duan}(2006{\natexlab{b}})}]{Yi2006a}%
  \BibitemOpen
  \bibfield  {author} {\bibinfo {author} {\bibfnamefont {W.}~\bibnamefont
  {Yi}}\ and\ \bibinfo {author} {\bibfnamefont {L.-M.}\ \bibnamefont {Duan}},\
  }\bibfield  {title} {\enquote {\bibinfo {title} {Detecting the breached-pair
  phase in a polarized ultracold {F}ermi gas},}\ }\href {\doibase
  10.1103/PhysRevLett.97.120401} {\bibfield  {journal} {\bibinfo  {journal}
  {Phys. Rev. Lett.}\ }\textbf {\bibinfo {volume} {97}},\ \bibinfo {pages}
  {120401} (\bibinfo {year} {2006}{\natexlab{b}})}\BibitemShut {NoStop}%
\bibitem [{\citenamefont {Bloch}\ \emph {et~al.}(2008)\citenamefont {Bloch},
  \citenamefont {Dalibard},\ and\ \citenamefont {Zwerger}}]{Bloch2008a}%
  \BibitemOpen
  \bibfield  {author} {\bibinfo {author} {\bibfnamefont {I.}~\bibnamefont
  {Bloch}}, \bibinfo {author} {\bibfnamefont {J.}~\bibnamefont {Dalibard}}, \
  and\ \bibinfo {author} {\bibfnamefont {W.}~\bibnamefont {Zwerger}},\
  }\bibfield  {title} {\enquote {\bibinfo {title} {Many-body physics with
  ultracold gases},}\ }\href {\doibase 10.1103/RevModPhys.80.885} {\bibfield
  {journal} {\bibinfo  {journal} {Rev. Mod. Phys.}\ }\textbf {\bibinfo {volume}
  {80}},\ \bibinfo {pages} {885} (\bibinfo {year} {2008})}\BibitemShut
  {NoStop}%
\bibitem [{\citenamefont {Levinsen}\ and\ \citenamefont
  {Parish}(2015)}]{Levinsen2015}%
  \BibitemOpen
  \bibfield  {author} {\bibinfo {author} {\bibfnamefont {J.}~\bibnamefont
  {Levinsen}}\ and\ \bibinfo {author} {\bibfnamefont {M.~M.}\ \bibnamefont
  {Parish}},\ }\bibfield  {title} {\enquote {\bibinfo {title} {Strongly
  interacting two-dimensional {Fermi} gases},}\ }in\ \href {\doibase
  10.1142/9789814667746_0001} {\emph {\bibinfo {booktitle} {Annual Review of
  Cold Atoms and Molecules}}},\ Vol.~\bibinfo {volume} {3},\ \bibinfo {editor}
  {edited by\ \bibinfo {editor} {\bibfnamefont {K.~W.}\ \bibnamefont
  {Madison}}, \bibinfo {editor} {\bibfnamefont {K.}~\bibnamefont {Bongs}},
  \bibinfo {editor} {\bibfnamefont {L.~D.}\ \bibnamefont {Carr}}, \bibinfo
  {editor} {\bibfnamefont {A.~M.}\ \bibnamefont {Rey}}, \ and\ \bibinfo
  {editor} {\bibfnamefont {H.}~\bibnamefont {Zhai}}}\ (\bibinfo  {publisher}
  {World Scientific},\ \bibinfo {address} {Singapore},\ \bibinfo {year}
  {2015})\ Chap.~\bibinfo {chapter} {1}, pp.\ \bibinfo {pages}
  {1--75}\BibitemShut {NoStop}%
\bibitem [{\citenamefont {Jeszenszki}\ \emph {et~al.}(2018)\citenamefont
  {Jeszenszki}, \citenamefont {Cherny},\ and\ \citenamefont
  {Brand}}]{Jeszenszki2018a}%
  \BibitemOpen
  \bibfield  {author} {\bibinfo {author} {\bibfnamefont {P.}~\bibnamefont
  {Jeszenszki}}, \bibinfo {author} {\bibfnamefont {A.~Yu.}\ \bibnamefont
  {Cherny}}, \ and\ \bibinfo {author} {\bibfnamefont {J.}~\bibnamefont
  {Brand}},\ }\bibfield  {title} {\enquote {\bibinfo {title} {\textit{s}-wave
  scattering length of a {G}aussian potential},}\ }\href {\doibase
  10.1103/PhysRevA.97.042708} {\bibfield  {journal} {\bibinfo  {journal} {Phys.
  Rev. A}\ }\textbf {\bibinfo {volume} {97}},\ \bibinfo {pages} {042708}
  (\bibinfo {year} {2018})}\BibitemShut {NoStop}%
\end{thebibliography}
%
%

\end{document}